\documentclass[fleqn,usenatbib]{mnras}

\usepackage{newtxtext,newtxmath}

\usepackage[T1]{fontenc}

\newcommand{\ith}{\ensuremath{^{\rm th}}}

\newcommand{\alphafe}{[$\alpha$/Fe]}

\newcommand{\ha}{high-$\alpha$}
\newcommand{\highalpha}{high-$\alpha$}
\newcommand{\lowalpha}{low-$\alpha$~}

\newcommand{\dr}{$R_{\rm now}- R_{\rm bir}$}
         
\DeclareRobustCommand{\VAN}[3]{#2}
\let\VANthebibliography\thebibliography
\def\thebibliography{\DeclareRobustCommand{\VAN}[3]{##3}\VANthebibliography}

\usepackage{graphicx}	
\usepackage{amsmath}	

\title[Turning Points in the Age-Metallicity Relations]{Turning Points in the Age-Metallicity Relations --- Created by Late Satellite Infall and Enhanced by Radial Migration}

\author[Lu et al.]{
Yuxi (Lucy) Lu,$^{1, 3}$\thanks{E-mail: lucylulu12311@gmail.com}
Melissa K. Ness,$^{1, 2}$
Tobias Buck$^{4}$, 
Christopher Carr$^{1}$
\\
$^{1}$Department of Astronomy, Columbia University, 550 West 120\ith\ Street, New York, NY, USA\\
$^{2}$Center for Computational Astrophysics, Flatiron Institute, 162 5\ith\ Avenue, Manhattan, NY, USA\\
$^{3}$American Museum of Natural History, Central Park West, Manhattan, NY, USA\\
$^{4}$Leibniz Institute for Astrophysics Potsdam, An der Sternwarte 16, 14482 Potsdam, Germany\\
}

\date{Accepted XXX. Received YYY; in original form ZZZ}

\pubyear{2021}

\begin{document}
\label{firstpage}
\pagerange{\pageref{firstpage}--\pageref{lastpage}}
\maketitle

\begin{abstract}
The present-day Age-Metallicity Relation (AMR) is a record of the star formation history of the Galaxy, as this traces the chemical enrichment of the gas over time. We use a zoomed-in cosmological simulation that reproduces key signatures of the Milky Way (MW), g2.79e12 from the NIHAO-UHD project, to examine how stellar migration and satellite infall shape the AMR across the disk. We find in the simulation, similar to the MW, the AMR in small spatial regions (R, z) shows turning points that connect changes in the direction of the relations. The turning points in the AMR in the simulation, are a signature of late satellite infall. This satellite infall has a mass radio similar as that of the Sagittarius dwarf to the MW ($\sim$ 0.001). Stars in the apex of the turning points are young and have nearly not migrated. The late satellite infall creates the turning points via depositing metal-poor gas in the disk, triggering star formation of stars in a narrow metallicity range compared to the overall AMR. The main effect of radial migration on the AMR turning points is to widen the metallicity range of the apex. This can happen when radial migration brings stars born from the infallen gas in other spatial bins, with slightly different metallicities, into the spatial bin of interest. These results indicate that it is possible that the passage of the Sagittarius dwarf galaxy played a role in creating the turning points that we see in the AMR in the Milky Way.
\end{abstract}




\section{Introduction} \label{sec:intro}
Large spectroscopic surveys such as the Apache Point Observatory Galactic Evolution Experiment (APOGEE) \citep{Majewski2017}, the Large Sky Area Multi-Object Fibre Spectroscopic Telescope (LAMOST) \citep{LAMOST}, the GALactic Archaeology with HERMES (GALAH) \citep{Silva2015, Buder2019}, Gaia-ESO \citep{Gilmore2012}, and the Radial Velocity Experiment (RAVE) \citep{RAVE1, RAVE2} have together observed millions of stars.
These provide a large number of spectroscopic stellar ages and detail abundances to study the formation and evolution of the Galaxy.

In our recent paper, we investigated the high- and \lowalpha disk stars of the Milky Way using APOGEE DR16, and found that the two disk populations exhibit some similar properties \citep{Lu2021}.
In particular, the intrinsic dispersion around age-abundance relations are, for most elements, very small. 
We also investigated the AMRs for both the high and low-$\alpha$ disk separately, expanding upon \citet{Feuillet2019}. 
We determined that whilst the two populations show different AMRs across Galactic radius and heights, both show structure, with bends and ``turning points''.
\cite{Lu2021} also pointed out that by comparing between simulations and analytic models with data, we could further constrain the formation scenarios for the MW. 
In this follow up paper, we investigate the shape of the Age-Metallicity Relation (AMR) as a potential fossil record of Milky Way's (hereafter MW) formation history by analyzing one of the MW-like cosmological zoom-in simulations from NIHAO-UHD \citep{Buck2020, Buck2020b}. 

The AMR was first observed in the solar neighborhood by \cite{Twarog1980}, revealing an overall decrease in metallicity with age. 
This work concluded that the AMR can be used to estimate the star formation rate, via comparison to theoretical models described in \cite{Larson1972}. 
This means that the AMR can provide important insight into the star formation history of galaxies.

The wide spread in stellar metallicity in the solar neighborhood, at any given age, suggests radial migration has taken place, as stars are believed to be born in a narrow metallicity distribution in within a small age and radius bin.
Radial migration is a process where stars move away from their birth locations over time.
This was first suggested by \cite{Grenon1972, Grenon1989}, in which they found old metal-rich stars in the solar neighborhood that have kinematic and abundance properties consistent with those of the inner disk.
There are two types of radial migration, churning and blurring. 
Churning describes a change in the angular momentum (or guiding center) of an orbit without an increase in orbital heating. Blurring, on the other hand, is a change in a star's orbital eccentricity without a change in the angular momentum.
\cite{Sellwood2002} suggested transient spiral modes can cause radial migration, mostly at the co-rotation resonance.
Other non-axisymmetric perturbations to the potential, such as bars or galaxy interactions, are also important mechanisms that can cause radial migration, as seen in simulations \citep[e.g.][Carr et al. in prep]{Bird2012, Roskar2008, Quillen2009}.
Churning can flatten the intrinsic metallicity gradient in the disk significantly overtime as the metal-rich stars from the inner galaxy migrate outward.
Blurring can also affect the apparent metallicity gradient as stars move away from their guiding radius on eccentric orbits \citep{Sellwood2014}.

The strength of radial migration has been estimated to be 3.9 kpc $\sqrt{\tau/7\ Gyr}$ for disk stars \citep{Frankel2019}.
\cite{Frankel2020} also concluded that the effect of blurring is one magnitude smaller than that of churning using an analytic model fitted to APOGEE data. 
However, in this paper, we do not distinguish between the two processes and instead, study the broader effect of radial migration on the AMR. 

One intriguing finding, described in \cite{Feuillet2018}, is the turnover points in the AMR after splitting stars into various spatial bins. 
They found by binning stars in metallicity, the youngest stars are not the most metal enriched stars throughout the disk as previously expected from gas enrichment processes.
This could either result from 1) two forms of star formation, where metal-rich gas is diluted, to form stars that are subsequently more metal-poor, or 2) migration of the old, metal-rich stars from the inner to the outer regions of the galaxy.

A direct approach for studying radial migration with data, is to infer the birth radii of the stars.
By combining age information with birth radii, one can study the process of radial migration and the formation history of the MW. 
However, birth radii is  not a direct observable.
One approach for inpainting stars with their birth radius is by combining age and metallicity, assuming a model for the overall time evolution of the ISM metallicity, using a reference radius such as solar vicinity, and the ISM metallicity gradient.
However, the empirical information on the temporal evolution of ISM metallicity ([Fe/H]$_{\rm ISM}(r, t)$) is mostly lost due to radial mixing, and thus, assuming a functional form of the ISM time evolution may not be the best approach \citep[e.g.][]{Minchev2018}.
Moreover, other external perturbations such as satellite infall or galaxy mergers can also prompt radial migration \citep[e.g.][Carr et al. in prep.]{Roskar2008} and further complicate the inferring process.
The most recent satellite interaction with the MW is believed to be Sagittarius within the pass billion year \citep[e.g.][]{Ibata1994, Laporte2018}. 

Another way to study radial migration in spiral galaxies, is to analyze simulations. 
In this way, the measurements that can be made from stars in the MW observational data sets can be compared and contrasted with the predicted trends seen in simulations -- in the context of their birth versus current day radius in the simulation.
In looking to use simulations to interpret observations in the MW, we want to be reassured that the simulations are broadly representative of the MW Galaxy.
Recent simulations \citep[e.g.][]{Aumer2013,Stinson2013,Marinacci2014,Wang2015,Grand2017,Hopkins2018,Buck2020b,Buck2019c} have successfully reproduced key observations in disk galaxies. 
Although feedback mechanisms and parameter selections can greatly affect the results \citep[e.g.][]{Keller2019,Dutton2019,Munshi2019,Buck2020b}, these simulations are able to provide important insights on the mechanisms that are in play during galaxy formation.

In this paper, we focus on the g2.79e12 simulation from the Numerical Investigation of a Hundred Astronomical Objects \citep[NIHAO][]{Wang2015} project. 
This galaxy\footnote{The simulation data is publicly available at\\ \url{{https://tobibu.github.io/\#sim_data}}.} is described in detail in \citep{Buck2020b} and has a spiral stellar disk with a strong bar \citep{Buck2018,Buck2019} and a well defined high- and \lowalpha disk, which closely resembles the MW \citep[][see also Figure 1]{Buck2020}. 
The simulation g2.79e12 is well suited for our study, as it has been previously used to study dynamical features of the bar \citep{Hilmi2020} and to compare metal poor stars to observational findings from the Pristine Survey \citep[][]{Sestito2021}.

We compare our simulation with data described in \cite{Lu2021} both from a global viewpoint (Section \ref{subsec:global}) and in the [O/Fe]-[Fe/H] plane (Section \ref{subsec:detail}). 
We describe the process via which we compare the simulations to the data in a consistent way in Section \ref{sec:datamethod}.
In Section \ref{subsec:feage}, we are able to reproduce the observed turning points in the AMR, with key features.
We also find late satellite infall events with a mass ratio on the order of Sagittarius to the MW is enough to trigger the creation of these turning points, and these turnovers can then be enhanced by radial migration.
Finally, in Section \ref{sec:dis}, we interpret some of the characteristics found in the AMR of the MW, under the assumption that the MW Galaxy follows a similar formation history as the simulation.

\section{Data \& Methods} \label{sec:datamethod}
\subsection{Data}\label{subsection:data}
{\it Observational data} --- Our observational data selection is inherited from \cite{Lu2021}. Here we selected $\sim$ 64,000 APOGEE DR16 \citep{Majewski2017} red giant stars and inferred their stellar ages using \texttt{The Cannon} \citep{Ness2015}.
We also used the \texttt{StarHorse} \citep{Queiroz2018} distances to select stars from the simulation in Section \ref{subsection:method}.

{\it Simulation data} --- The simulated galaxy studied in this work was first presented by \citep{Buck2018}. It is from a suite of high-resolution cosmological hydrodynamical simulations of MW-mass galaxies from the NIHAO-UHD project \citep{Buck2020b}. 
This galaxy model was calculated using a modified version of the smoothed particle hydrodynamics (SPH) solver GASOLINE2 \citep{Wadsley2017}. 
The simulation is calculated from cosmological initial conditions and star formation and feedback are modelled following the prescriptions in \citep{Stinson2006,Stinson2013}. 
The total stellar mass is $1.59 \times 10^{11} M_\odot$. 
The galaxy is resolved with $\sim8.2\times10^6$ star, $\sim2.2\times10^6$ gas, and $\sim5.4\times10^6$ dark matter particles \citep[see Table 1 in][]{Buck2020b}.  This corresponds to a baryonic mass resolution of $\sim3\times 10^4 M_\odot$ per star particle ($\sim9 \times 10^4 M_\odot$ gas particle mass) or 265 pc force softening. 

During the run-time of the simulation over $\sim$ 14 Gyrs, we follow the growth of structures from cosmological initial conditions, star formation, gas accretion and successive mergers of dwarf galaxies, with the central disk galaxy. 
The final central disk galaxy is surrounded by dozens of satellite galaxies which closely follow the observed satellite mass function of either the MW or M31 \citep{Buck2019b}.
Most importantly for this study, we focus on the simulation g2.79e12, since it shows a clear bimodality in the [Fe/H]-[$\alpha$/Fe] abundance plane, in the disk \citep{Buck2020}. 
For this particular simulation, we are interested in understanding if and how the satellite infall event that started around z = 0.34 ($\sim$ 4 Gyr from the present day) might affect the AMR.
For more information on the simulation details and galaxy properties we refer the reader to \cite{Buck2020b}. 

\subsection{Methods}\label{subsection:method}
We first select stars from the simulation g2.79e12 that have birth radii, $R_{\rm bir}$ $<$  15 kpc and absolute birth height, $z_{\rm bir}$ $<$ 5 kpc. This ensures the stars we study are born in the disk, and not in accreted satellites.
This reduces our sample size from $\sim$ 8 million to $\sim$ 7.1 million star particles.

\begin{figure*}[htp]
    \centering
    \includegraphics[width=\textwidth]{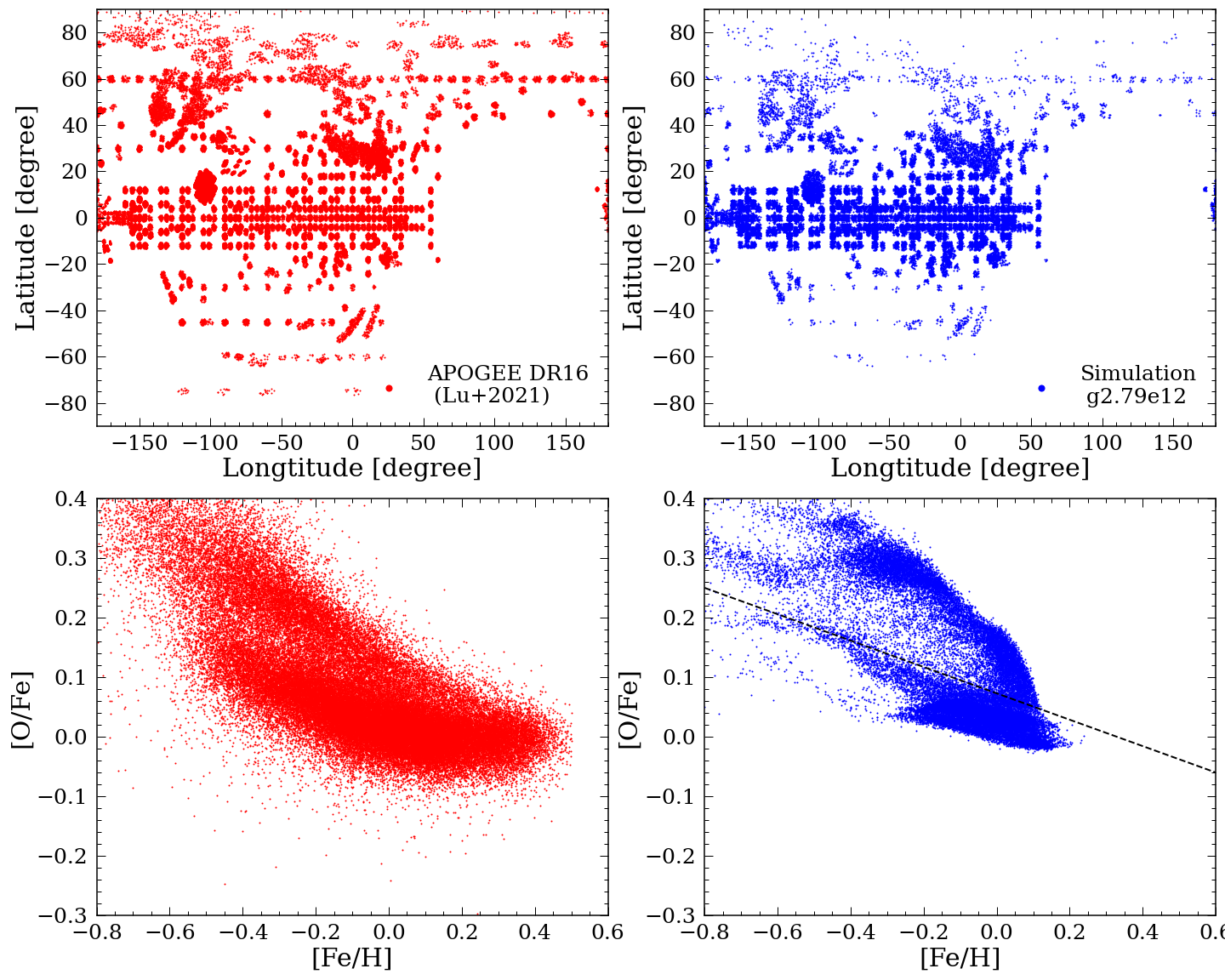}
    \caption{Our sample of 35,944 stars selected from the g2.79e12 simulation \citep{Buck2020b} from the APOGEE sampling on the sky (right two plots; blue) and data from Lu et al. (2021) (left two plots; red).
    This selection function is only used while comparing our simulation results to data in Lu et al. (2021) in section~\ref{subsec:global} and section~\ref{subsec:detail}. 
    Due to the resolution of the simulation, we are not able to find matching stars in the simulation for some stars in the data.}
    \label{fig:select}
\end{figure*}

In order to match the simulation with APOGEE data, we grid the APOGEE data into 300$\times$150 squares in galactic longitude and latitude, respectively. 
We then select star particles that are closest to the observed stars, which are also within the range of distances presented in the data, if available.
Figure~\ref{fig:select} shows the selected 35,944 stars from the g2.79e12 simulation \citep{Buck2020b} in blue, and the 64,317 stars from \cite{Lu2021} in red.
Due to the resolution of the simulation, we are not able to match every single star in the data with one in the simulation.
Note that this is not the most sophisticated method to account for the APOGEE selection function since each star particle (average mass $\sim$ 20,000 $M_\odot$) in the simulation represents one group of stars (for a more strategic method, see \cite{Valentini2018}).
One other issue with this method is that although the scale heights of these simulations are similar to the MW, the scale length (5.57 kpc in simulations) are up to 2 times larger than that of the MW ($\sim$ 3 kpc \citep{Bland2016}) \citep{Buck2020b}. 

The top plots show the spatial selection of the stars and the bottom plots show the [Fe/H]-[O/Fe] plane for the data (left) and simulation (right). 
The black dashed line is an ad-hoc separation of the high- and \lowalpha disk for the simulation.

After selecting the simulation stars using our spatial transformation of the APOGEE selection, we scale the [O/Fe] and [Fe/H] by first centering the mean with the mean of the data in these two abundances.
We then multiply [O/Fe] and [Fe/H] of the simulation by the ratio of the standard deviation between data and the simulation.
Although the high- and \lowalpha disks can be considered as not two discrete populations \citep[e.g.][]{Bovy2012,Lu2021}, we divided the simulation into two disks for the purpose of comparing simulation with the results in \cite{Lu2021}.

We note that by re-analyzing the simulation without matching the APOGEE spatial selection, we draw the same conclusions that are presented in Section \ref{sec:results}.
Therefore, we conclude, the method applied here is adequate for our purposes in this paper.

\section{Results}\label{sec:results}

The g2.79e12 simulation shows qualitative similarities to the MW. It shows a bimodality in the [Fe/H]-\alphafe\ plane, where the \lowalpha disk is produced by a merger event $\sim$ 10 Gyr from the present day. The simulation has an age distribution that shows on average older stars in the inner galaxy and younger stars in the outer from inside-out formation. Additionally, flaring in the \lowalpha disk, features in the metallicity skewness and the AMR are shared between the simulation and Milky Way. 

We first compare the global age distribution as well as the metallicity skew with data, in Section \ref{subsec:global}.
We then split the simulation into chemical cells following the method described in \cite{Lu2021} in the \alphafe-[Fe/H] plane in Section \ref{subsec:detail}.
In Section \ref{subsec:feage}, we calculate the AMR using the full simulation and study how both a late merger (of a system with a mass similar to Sagittarius) and radial migration can create the turnover points.

By comparing the age and metallicity skew distributions between the data and simulation, we find very similar properties.
Figure 14 in \cite{Buck2020b} quantifies the role of heating and upside-down formation in this simulation.
We are able to validate that the lack of metallicity skewness gradient over radius does not mean that there is no radial migration present in the \ha~disk (see Sections \ref{subsec:global} and \ref{subsec:detail}).

\subsection{Global age distribution and metallicty skew}\label{subsec:global}
\begin{figure*}[htp]
    \centering
    \includegraphics[width=0.9\textwidth]{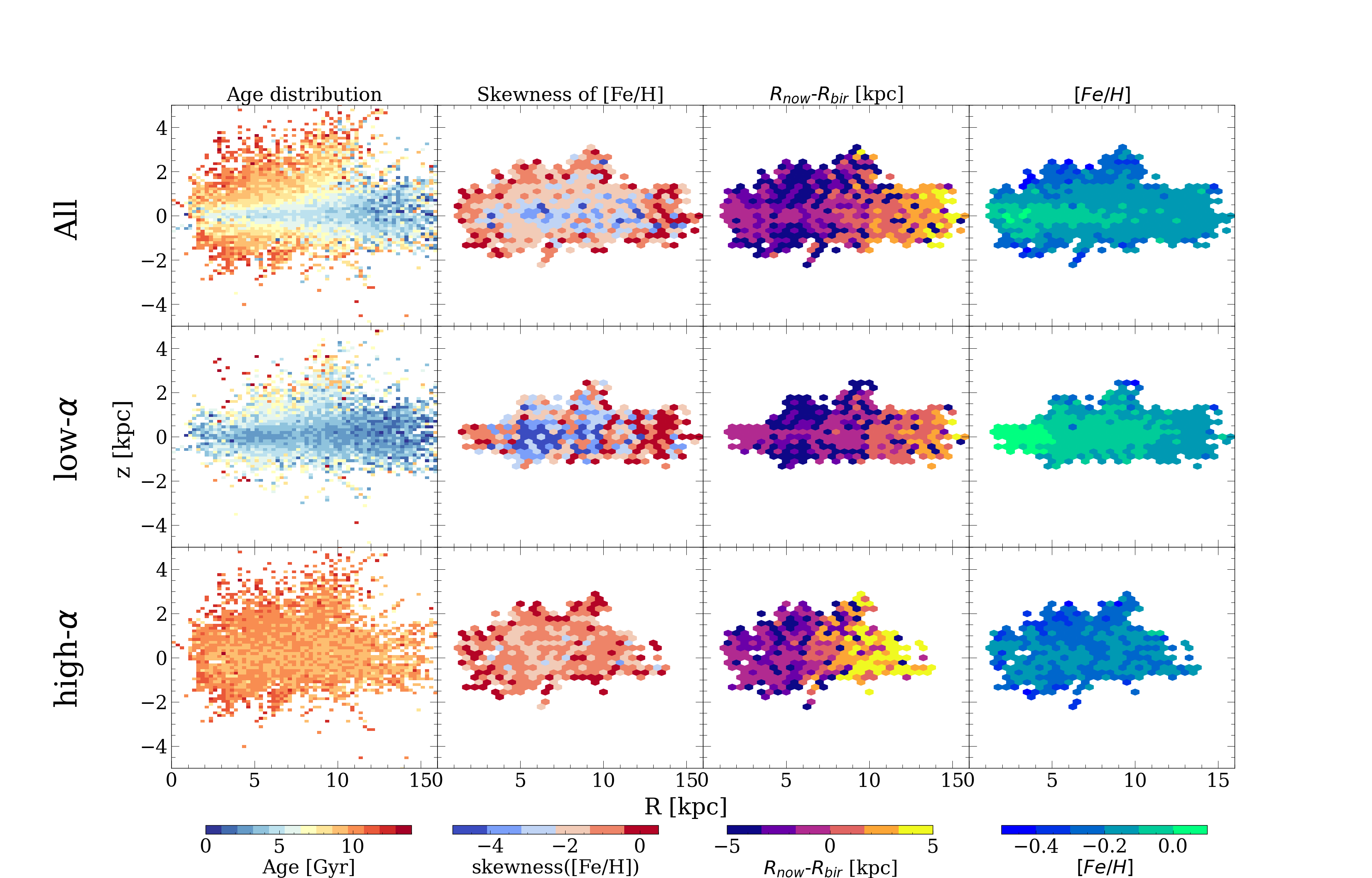}
    \caption{Age distribution (left), metallicity skew (second column), radial migration (third column), and metallicity distribution (right) for the entire sample (top), the \lowalpha disk (middle), and the \ha\ disk (bottom). Bins with fewer than 10 stars are masked in the metallicity skewness plots. We see very similar age distributions between simulation and data from Lu et al. (2021).The metallicity skewness also shows a slight gradient in the \lowalpha disk, and no gradient in the \ha\ disk. It is clear that radial migration is present in both disks, and the lack of skewness gradient in the \ha\ disk is not a signature of no radial migration. However, the metallicity skewness is always almost negative in the simulation, as compared to a negative to positive transition with increasing radius as the data Lu et al. (2021).}
    \label{fig:global}
\end{figure*}

In Figure~\ref{fig:global}, we compare the simulation (after applying the spatial selection) with Figure 5 from \cite{Lu2021}.
The first column shows the age distribution, the second column shows the metallicity skewness, the third column shows the change in radius ($R_{\rm now}- R_{\rm bir}$), and the last column shows the metallicity distribution.
The rows represent the distributions for all stars (top), \lowalpha  stars (middle), and \ha~ stars (bottom).

We list a series of similarities and differences and their interpretations: 

{\it Similarities in age distribution for both disks} --- We find the \lowalpha disk is young and the \ha\ disk is old. 
Similar to the data, the \lowalpha disk in the simulation flares up towards the outer galaxy, with the youngest stars centered in the mid-plane and the outer disk.
In contrary to the \lowalpha disk, the \ha\ disk does not have a clear age gradient across radius. 
However, the oldest stars in the simulation, similarly to data, are mostly located at high Galactic heights, indicating an upside-down formation history or heating of the old population.

{\it Similarities in the metallicity skewness} --- 
The skewness is used as a measurement of how far the distribution deviates from a Gaussian in each bin. 
A positive metallicity skew means there is a population of stars with metallicity higher than the average [Fe/H] in that bin, and visa-versa.  

\cite{Hayden2015} and \cite{Loebman2016} suggest the change in the direction of the metallicity skewness across radius in the disk as a possible signature of radial migration.

We find a slight gradient in metallicity skewness for the \lowalpha disk, where the skewness becomes more positive towards the outer disk, indicating radial migration.

We also find no clear skewness gradient in the \ha~disk. 
From the $R_{\rm now}- R_{\rm bir}$ plot, both disks experience radial migration, with the \ha\ disk stars having migrated farther in the outer disk, compared to the stars in the \lowalpha disk.
As a result, we can see that the absence of any skewness in the [Fe/H] with galactic radius, does not correspond to no radial migration \citep{Lu2021}. 

{\it Discrepancy in the \lowalpha disk metallicity skew gradient} --- The metallicity skewness gradient in the \lowalpha disk is not as strong as that in the data. 
However, as mentioned in \cite{Buck2020}, instead of the high- and \lowalpha disks, the g2.79e12 simulation shows three sequences.
If we select stars that are the most \alphafe-poor, the gradient becomes stronger, which matches closer with what we observe in the MW.
The weaker gradient could also arise if the radial migration is stronger in the MW than in the simulation. 

{\it Discrepancy in the overall \ha\ disk metallicity skewness} --- The metallicity skewness is almost exclusively negative for the \ha\ disk, instead of positive as shown in the data. 
After investigating further, this could be caused by our definition of high- and \lowalpha disk as the g2.79e12 simulation has three sequences instead of two.
By selecting the highest \alphafe~ sequence (\alphafe~$>$ 0.2 in Figure~\ref{fig:select}), we were able to measure a more positive metallicity skew.
Other factors could include uncertainties on the data or the differences in the MW formation history.

\subsection{Age distribution in the chemical plane}\label{subsec:detail}
\begin{figure*}[htp]
    \centering
    \includegraphics[width=\textwidth]{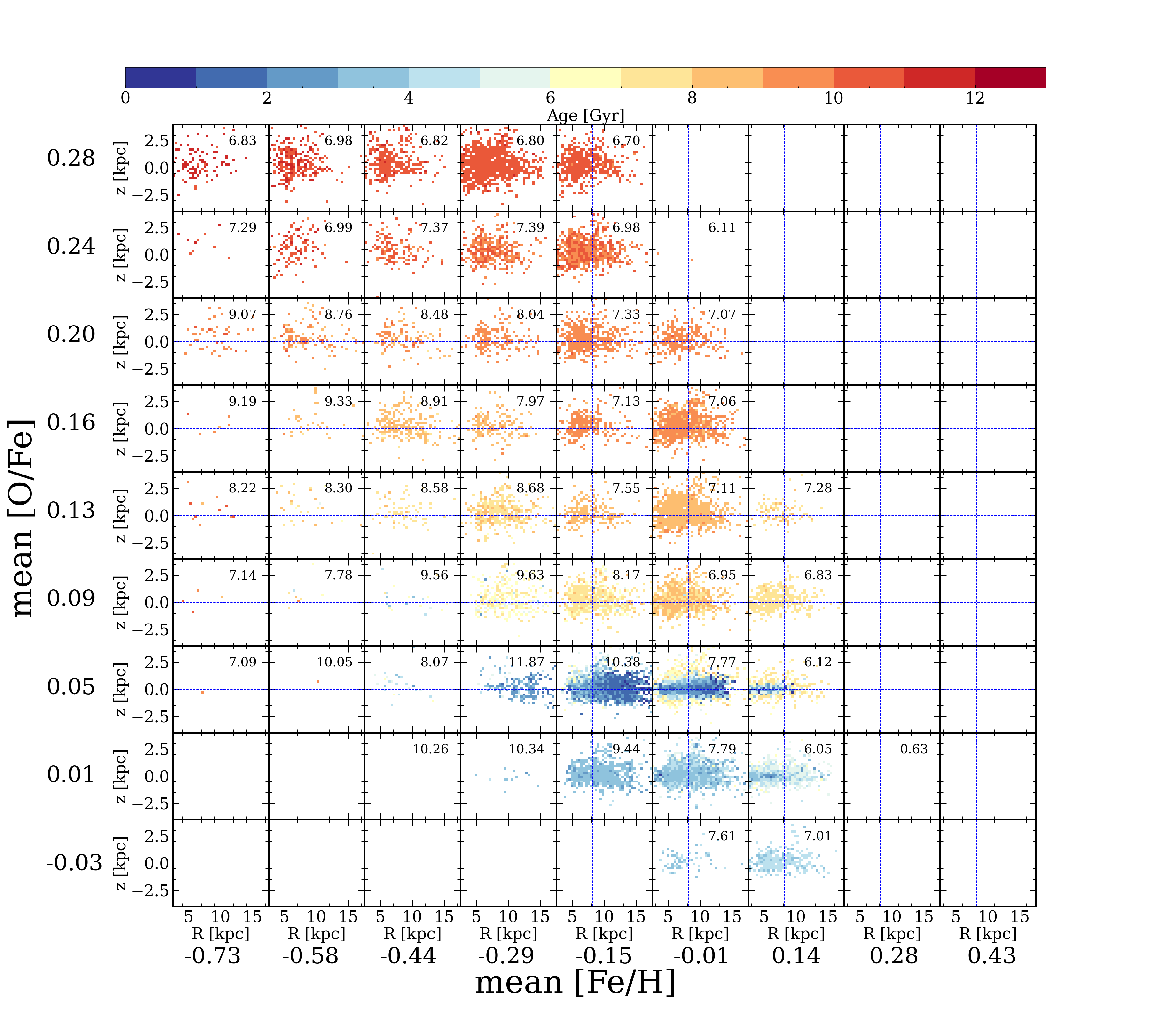}
    \caption{The spatial age distribution for the g2.79e12 simulation after applying the spatial selection function. 
    The mean values of [O/Fe] and metallicity for each subplot are displayed along the axis, and the bin sizes are 0.05 dex and 0.12 for [O/Fe] and metallicity, respectively.
    The blue dashed lines mark the disk mid-plane and solar radius. }
    \label{fig:cc}
\end{figure*}

Figure~\ref{fig:cc} shows the chemical cell plot, across bins in [Fe/H]-[O/Fe] for the simulation.
The mean [Fe/H] in each cell increases along the $x$-axis, and the mean [O/Fe] increases along the $y$-axis.
Each subplot shows the spatial distribution for stars within a small range of metallicity, [Fe/H] and [O/Fe], colored by age.
Each cell, or bin, is 0.05 dex wide in [O/Fe] and 0.12 dex wide in [Fe/H], and the spatial age distribution is plotted in each.
The blue dashed lines mark the disk mid-plane and the solar radius.
Compared to Figure 7 from \cite{Lu2021}, the simulation in the chemical plane shows a number of similarities with the data:

The age and spatial distributions of the stars in each chemical cell are similar to what is seen in the data, with the \ha\ disk being old and the \lowalpha disk young.
The most metal enriched \ha\ stars (top right corner) have smaller mean stellar ages and are more spread out radially compared to the metal-poor \ha\ stars (top left corner), indicating inside-out formation of the \ha\ disk. We also find that stars are more concentrated to the mid-plane, evolving from the \ha~disk to the \lowalpha disk (from top to bottom).

There is evidence that old \lowalpha and young \ha\ stars exist in the MW \citep[e.g.][]{Martig2015, Chiappini2015, Feuillet2018, Hekker2019, Lu2021}. However, we do not see these stars in the simulation.
This could be caused by 1) wrong stellar ages assigned to stars in observations, or 2) missing physics in the simulation, or both.
For 1), outliers that have abnormal surface properties (e.g. abundances, surface temperature, surface gravity) could lead to inaccurate inferred stellar ages.
Moreover, stellar age is not a direct observable but an estimation of a star's evolutionary stage \citep{Soderblom2010}.
This means, any stars that went through mergers or any accretion events can appear younger than they actually are. 
Studies have shown that the young \ha~stars are likely old stars that have experienced mass transfer and subsequently only appear to be young when they are in fact old  \citep[e.g.][]{Zhang2021,Yong2016,Jofre2016,Sun2020}.
Nonetheless, observational uncertainties are most likely not the sole reason for these old and young outliers.
For 2), other formation scenarios such as clumpy formation \citep[e.g.][]{Clarke2019} or the two in-fall model \citep[e.g.][]{Spitoni2020} could possibly reproduce old/young low-/\ha\ stars.
The lack of these stars in the NIHAO-UHD simulations, could indicate the lack of physics that allow local self-enrichment.

\subsection{Age-Metallicity relations and turning points --- collective effects of radial migration and late mergers}\label{subsec:feage}
\begin{figure*}[htp]
    \centering
    \includegraphics[width=\textwidth]{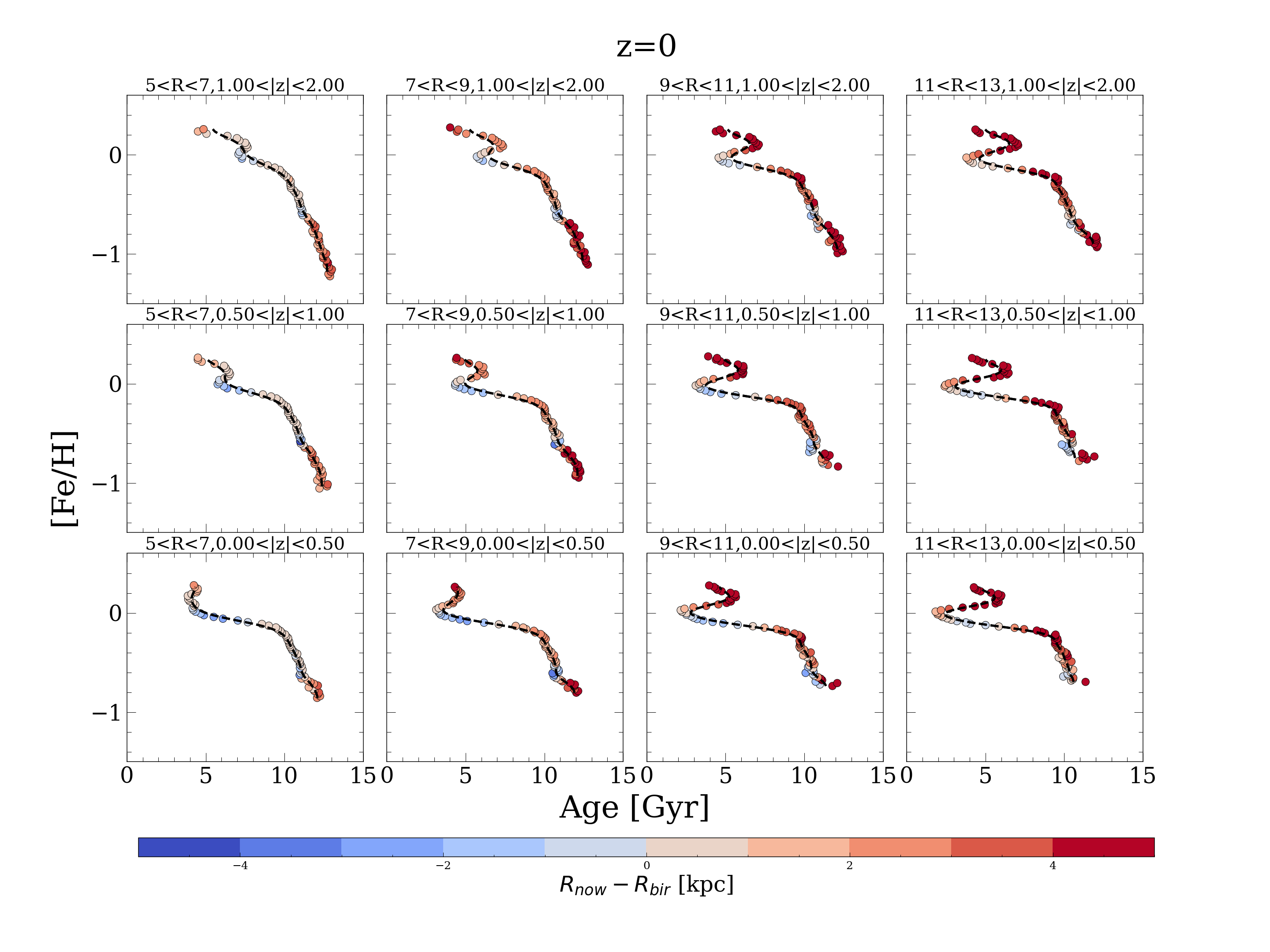}
    \caption{The AMR in different spatial bins across $R-z$; $R$ increases towards the right and galactic height, $z$ increases towards the top. 
    The points are colored by $R_{\rm now}- R_{\rm bir}$, showing the migration strength and direction (red indicates stars moved outwards to their current location), the average number of stars in each bin is $\sim$ 700. 
    Figure~\ref{fig:fehz0_hilow} shows the same figure but broken up by the high- and \lowalpha disks.}
    \label{fig:feagez0}
\end{figure*}

To study the AMR and the correlation between radial migration and the turning points, we utilize the full simulation without down-sampling to increase the sample size of stars. 
We confirm, using a sub-sampled simulation, that the survey-matched selection merely increases the scatter in the AMR, but does not change any of the overall results we report. The turning points in the AMR are the locations where the relation deviates from a function that increases in [Fe/H] with decreasing age. Starting from the oldest and most metal-poor stars, the mean age of stars decreases with increasing [Fe/H]. This reaches an apex of youngest stars, and then turns such that the age then increases with increasing [Fe/H]. This turning point in the AMR is seen in each panel in Figure~\ref{fig:feagez0}. 
As mentioned in Section \ref{sec:intro}, the turning points in the AMR are believed to be one major signature of radial migration \citep[e.g.][]{Feuillet2018,Lu2021}.
The logic behind this being that the metallicity decreases with increasing age below the turning point results from the natural enrichment of gas.
The metallicity increases with age above the turning point is a imprint of older, metal enriched stars migrating from the inner galaxy. 

We first investigate the correlation between radial migration and the turning points by reproducing Figure 3 from \cite{Feuillet2018} with the simulation.
Figure~\ref{fig:feagez0} shows the result from the simulation for four different radius bins (columns) and three different galactic heights (rows). 
Points are colored by $R_{\rm now}- R_{\rm bir}$, and the positive numbers (red) indicate the stars in that bin mostly migrated outwards to the current location. 
One global feature is that more stars migrated outwards than inwards.
This is expected as the surface density of the disk decreases towards the outer disk, and thus the probability for a net-outward migration is larger. 

Looking at the smallest radius bin (5 kpc $<$ R $<$ 7 kpc), most of the stars  migrated less than R $<$ 2 kpc to reach the current location (5 kpc $<$ R $<$ 7 kpc).
The small migration distances in the inner galaxy are not unexpected, since the migration direction is mostly from the inner disk to the outer disk, and the maximum outward migration for stars in this radius bin can only be between 5 to 7 kpc.
The net-outward migration is caused by the mass profile of the galaxy, as more stars are concentrated in the center.
The migration distance increases towards the outer disk as stars can travel further to get to the outer disk from the inner galaxy.
Galactic height does not seem to impact the overall radial migration distance.

In the simulation, we are able to observe the turning points that are also found in the MW data. 
Similar to the data, the turning points shift to lower metallicity towards the outer disk (compare left most and right most panels).
Since stars at the turning points are the youngest stars without enough time to migrate, the shift in turning points represents the decrease in metallicity with the present-day radius. 
The prominence of the turning points also decreases with galactic height (compare upper and lower panels). 
This has been proposed to be due to either the lack of young stars, or the decrease in efficiency of radial migration at large galactic heights \citep{Feuillet2018}.
Later in this Section, we explain that the decline in turning point strength with height from the galactic plane is also a natural characteristic if the turning points are produced from star formation triggered by a late merger. 

In the simulation, we are able to answer whether the turnover points in the AMR, as seen in Figure~\ref{fig:feagez0}, are correlated with radial migration.
From Figure~\ref{fig:feagez0}, the mean migration, at each point on the AMR, shows that the stars along the turning points have migrated the least. This is not  unexpected, since they are among the youngest stars, that have had the least time to migrate.
In order to rule out the case that there are almost equal numbers of stars moving in and out around the turning points, we calculated the absolute value of $R_{\rm now}- R_{\rm bir}$ and found that the stars indeed did not move significantly.

In the MW, the most metal-rich stars above the turning points in the observed AMR are believed to be migrated stars from the inner Galaxy.
From the simulation, these stars indeed have the most positive $R_{\rm now}- R_{\rm bir}$ values (meaning that they were born at smaller radii than they currently preside). 
However, stars that have metallicity just below the turning points also have (on average) migrated outwards at a similar amount compared to the most metal enriched stars. 
Figure~\ref{fig:fehz0_hilow} shows how the migration direction is also, in detail, highly correlated with membership in the high or low-alpha sequence, as discussed in later Sections.

Another observation is that the turning points seem to be disruptions from a natural and smoothly evolving gas enrichment, meaning that if we excluded the turning points, they would be near-monotonic sequences.
These two observations (the stars above and below the turning points migrated the same amount, and that the turning points seem to be disruptions from the ``background'' AMR) suggest that stars at the turning points could be formed in a separate, distinct process, and not necessarily created via a continuous process such as radial migration alone. 
In Figure~\ref{fig:fehz0_nomig}, the AMR for stars that are born within each spatial bin, equivalent to examining the AMR without migration between bins, shows strong turning points in the AMR. 

One process that can disrupt the AMR of the disk is the accretion of a gas rich satellite, at a later epoch in the formation history of the galaxy. 
Such a merger event can bring in metal-poor gas,  forming young stars in the disk at a lower metalicity than before the accretion. This process can subsequently effectively generate a second AMR, that is offset from the existing one. Thereby, creating excursions in the AMR relation observed in the present day, of the turning points.

\begin{figure*}[htp]
    \centering
    \includegraphics[width=\textwidth]{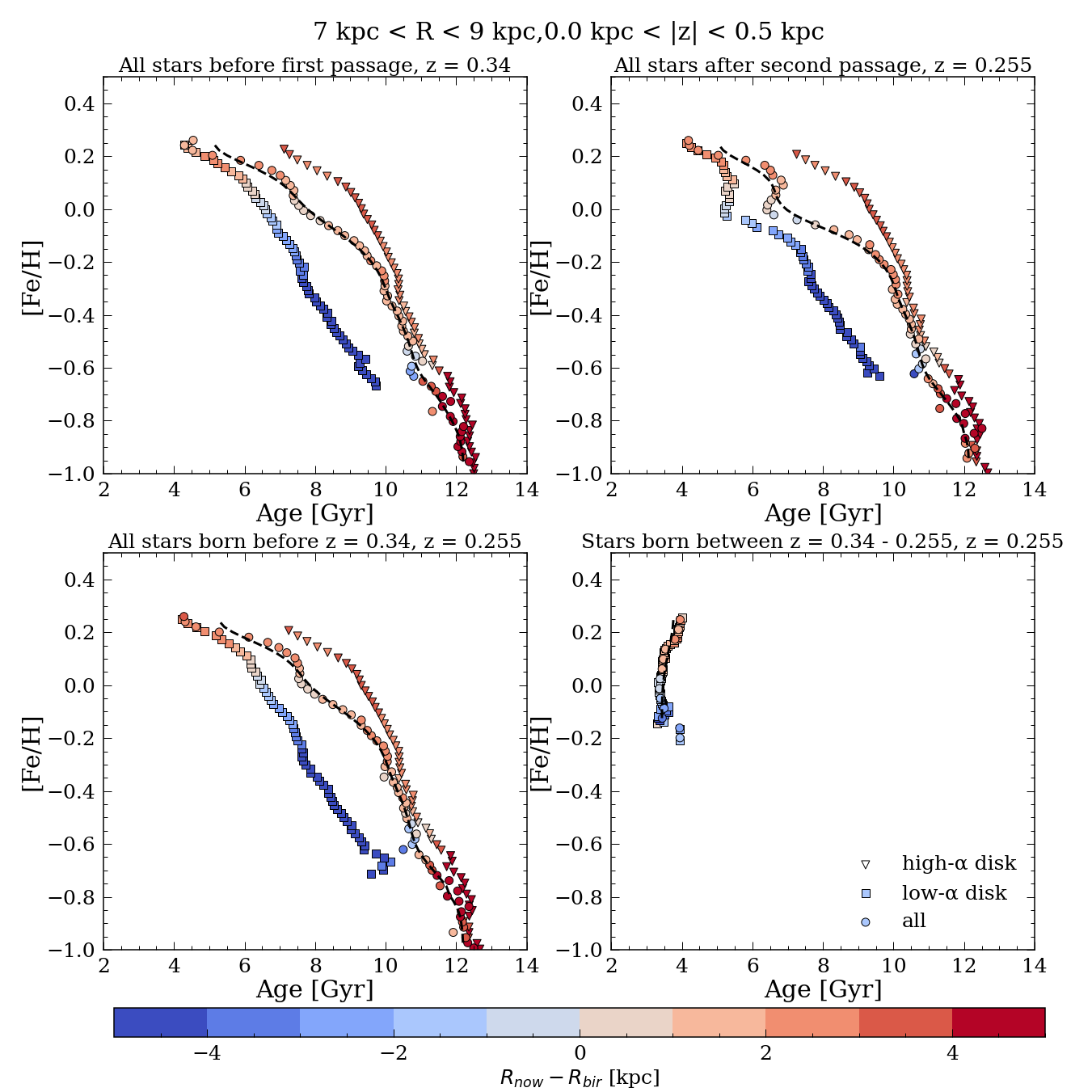}
    \caption{The effect of a satellite infall (mass ratio 0.001) on the AMR at 7 kpc $< R <$ 9 kpc and $|z| < $0.5 kpc). 
    The top left panel shows the AMR right before the first satellite passage, the top right plot shows the AMR right after the second satellite passage, the bottom left panel shows the stars that are born before the first passage at the time of the second passage, and the bottom right panel shows the stars that are created after the first passage.
    The first and second passage are $\sim$ 0.8 Gyr apart.
    Circles represent the AMR for all the stars, upside down triangles represnt that for only the \ha\ disk stars, and squares represent the AMR for only the \lowalpha disk stars.
    The points are colored by \dr.
    The bottom left plot shows the effect of radial migration, and the top right plot shows the collective effect of radial migration and a late major merger. }
    \label{fig:feagez034}
\end{figure*}

In the simulation used in this project, a satellite with a mass ratio of 0.001 started merging with the galaxy at redshift $\sim$ 0.34 (second passage at z $\sim$ 0.25), and we believe this infall is responsible for the creation of the turnover points. 
This process is shown in Figure~\ref{fig:feagez034}.
The top two panels of this Figure show the AMR before the first passage of the satellite (top left panel), and after the second passage of the satellite (top right panel).
The bottom left panel of Figure 5 shows the AMR after the second passage, but only for stars that were already born before the first passage -- thus isolating the effect of radial migration alone\footnote{We emphasize that the top left and the top right two panels are not necessarily identical. This is because the stars are free to migrate in and out of the radial bin (7 $<$ R $<$ 9 kpc).} The bottom right panel of Figure 5 shows the stars that are born after the first passage -- thereby showing the impact of the satellite infall in the AMR plane.
The three markers in each figure represent all the stars (circles), the \ha\ disk stars (upside down triangles), and the \lowalpha disk stars (squares).
The points are colored by \dr.
The first and second passages are $\sim$ 0.8 Gyr apart.

Looking at the overall AMR (circles) in the top left panel of Figure~\ref{fig:feagez034}, there is only a weak turning point before the first passage. This weak turning point originates from the difference in slope between the AMR of \lowalpha and \ha~stars in the metallicity range $-0.2 \lesssim$ [Fe/H] $\lesssim -0.1$.

Comparing the top left to the bottom left panels in Figure~\ref{fig:feagez034}, which show the stars born before z=0.34 at z=0.34 and z=0.255 respectively, we see that the turnover points do not change significantly. This suggests radial migration alone will not create the turnover points, as we are concerned here only with stars that exist before the merger.
The effect of the merger is shown in the bottom right panel of Figure~\ref{fig:feagez034}, where we show the AMR for stars that are born in the disk during the satellite infall event, as these stars are created from the metal-poor, \lowalpha gas that is brought in by the merger. The stars born in the merger event are co-eval.  
Furthermore, they are all younger than the other stars in the disk and have not migrated much (on average $\sim$ 0.5 kpc).  They also have a narrow range of metallicity ($-0.2$ dex $<$ [Fe/H] $<$ 0.2 dex), with more metal-rich stars born at the inner radial bin edge and more metal poor stars at the outer radial bin edge, compared to the rest of the stars that were born before the infall.

In order to validate that the stars born between the two passages of the satellite are indeed associated with the infall, we examine the age-[Fe/H] relation for stars born within small time intervals. 
We want to check if the satellite disrupts the slope of the age-[Fe/H] relation in the disk. 
We measure the mean metallicity, [Fe/H], and standard deviation of the [Fe/H], for the stars born between every two consecutive snapshots ($\sim$ 200 Myr apart) near the time of the first passage (see Figure~\ref{fig:fehsnap}).
By doing this, we see that the average metallicity, [Fe/H] for the newly born stars decreases right after the first satellite passage. 
This indicates that the turnover feature in the present-day AMR is associated with the infall event. 

After the second passage of the satellite (right top panel of Figure~\ref{fig:feagez034})), the major turnover point appears at $\sim$ 5 Gyr and [Fe/H] = 0.0 dex. 
The pronounced turning point here thereby mostly originates from new stars formed, as triggered by the satellite infall, around redshift z = 0.34, that brings metal-poor gas into the disk. 
There is no star formation in the \ha~disk from this infall, thus the \ha~AMR doesn't change.

The infall triggered AMR (bottom right panel of Figure~\ref{fig:feagez034}) combined with the existing one in the disk (bottom left panel of Figure~\ref{fig:feagez034}), shifts the average age in each [Fe/H] bin to lower values over a narrow range in [Fe/H] where the new stars are formed. The new stars (born in a narrow metallicity range) are significantly younger than all other stars in this [Fe/H] range at this galactic radius bin (i.e. compare the bottom two panels of Figure~\ref{fig:feagez034}). However, in order to create a turnover point with merger/satellite infall, enough metal-poor gas needs to be brought in by the merger. 
If a merger does not provide enough pristine low-metallicity gas, the self-enrichment process will quickly re-enrich the gas and bring the metallicity of the ISM back to the pre-merger level (see also the discussion for Fig.~\ref{fig:feage_ism}). Turnover points are also more likely to be created at a later time, after the frequent merger events have subsided.
In the early phase of galaxy evolution, when mergers are believed to happen frequently \citep[e.g.][]{Sestito2021}, any perturbations, e.g. from radial migration or single merger events, will be washed out by the short enrichment timescale of the gas. 

After examining the overall AMR, we now look more closely at the high- and \lowalpha stars separately. 
Signatures from both the radial migration and the merger are mostly observed in the \lowalpha disk. 
There are two reasons for this:
Studies suggest that the radial migration strength should be lower in the \ha~disk (or thick disk) as these stars have higher vertical actions, and are more resistant to radial migration \citep[e.g.][]{VeraCiro2014, Daniel2018, Minchev2018, Sellwood2002}.
However, other studies have shown that the decrease is mild \citep[e.g.][]{Solway2012, Kordopatis2015}, which is consistant with what we see here.
The merger signal is characterized by the dynamical perturbation of the already existing stars, as well as the new stars forming from the metal-poor gas that is brought in from the merger (bottom right).
Necessarily, the stars born during the merger event are \lowalpha stars and thus, formally the signature is only found for those stars.

\begin{figure*}[htp]
    \centering
    \includegraphics[width=\textwidth]{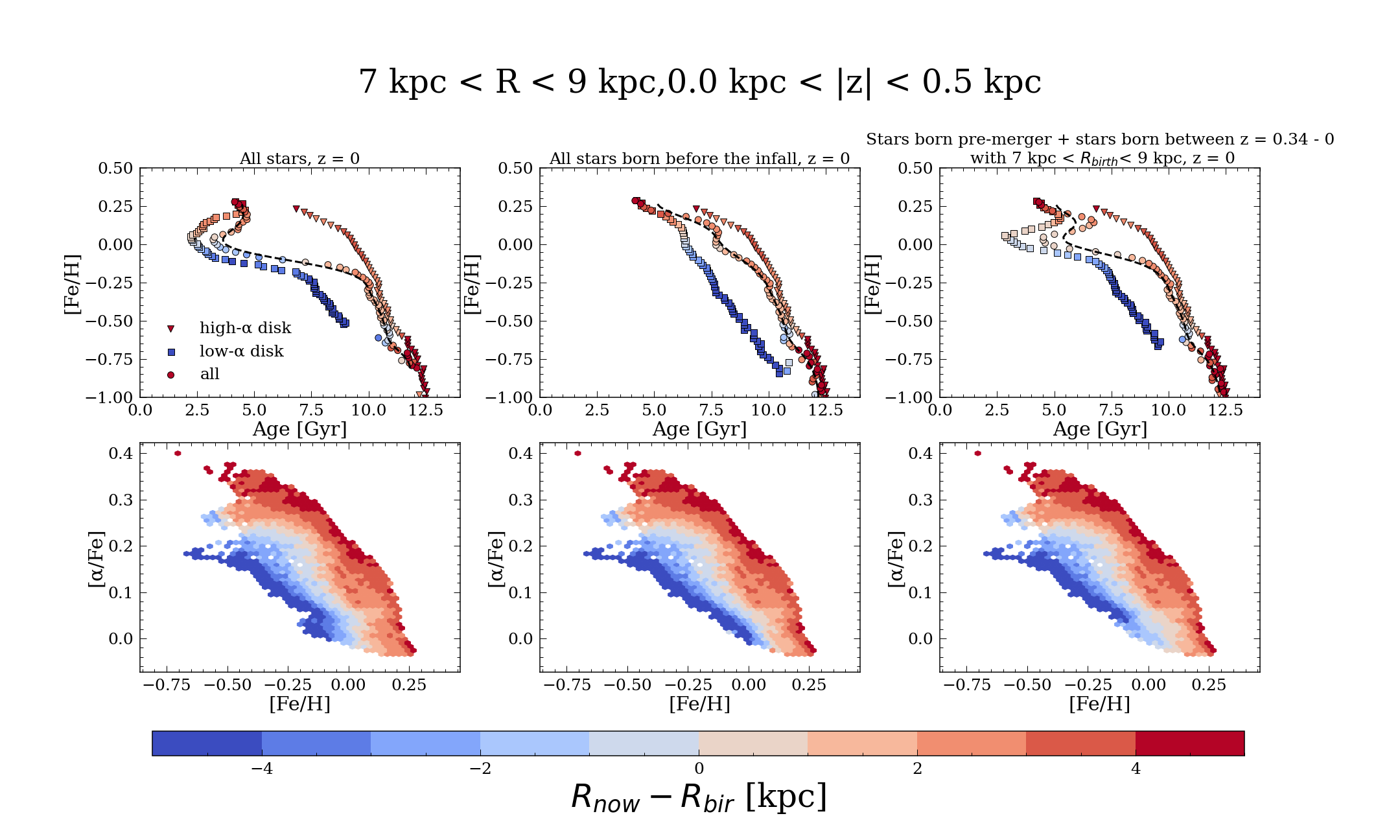}
    \caption{This Figure describes the effect of radial migration after the satellite infall created the turning points. 
    The top left plot shows the AMR for all the stars at redshift 0; the top middle plot shows the stars born before the merger (z = 0.34) at redshift 0; and the top right plot shows the stars born before the merger + stars born in the spatial bin of 7 $<$ R $<$ 9 kpc, $|z| <$ 0.5 kpc, between redshift 0.34 and 0. 
    The bottom plots show the stars above in [Fe/H]-\alphafe~plane.
    The top middle plot suggests that without the satellite infall bringing metal-poor gas, radial migration alone would not be able to create a strong turning point.
    However, the effect of radial migration after a merger is important. 
    This can be seen in the difference between the top left and the top right panels, as most stars excluded from the right panel are stars that are born from the metal-poor gas brought in by the merger and are {\it not} created in the 7 kpc $<$ R $<$ 9 kpc radius bin.
    These stars migrated from other radial bins to the current radial bin, thereby, enhancing (moving the apex of the turning point to a younger age) and widening the turning point along the [Fe/H] axis. }
    \label{fig:feagez0merger}
\end{figure*}

\begin{figure*}[htp]
    \centering
    \includegraphics[width=\textwidth]{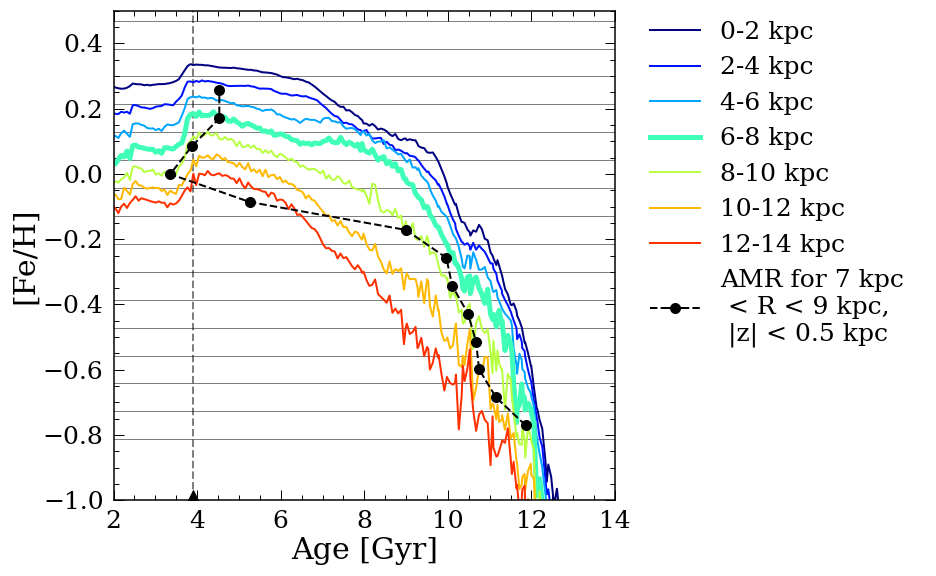}
    \caption{The colored lines show the ISM evolution over time at different radii taken from Buck (2019). The black dots show the AMR for stars at 7 kpc $<$ R $<$ 9 kpc and $|z| <$ 0.5 kpc.
    The errorbars (can be ignored in most metallicity bins) are calculated using the standard deviation of the stellar ages in each [Fe/H] bins.
    The triangle marker and the grey dashed line at $\sim$ 4 Gyr shows the time of the merger at the satellite's first passage. 
    The ISM evolution that is closest to the radial bin is shown in bold.
    The horizontal lines show the bin edges in metallicity.}
    \label{fig:feage_ism}
\end{figure*}

Notice that the turning point in the AMR around the solar neighborhood (see Figure~\ref{fig:feagez0}) is stronger and wider than that shown at redshift z = 0.255 (see Figure~\ref{fig:feagez034}). 
This is a collective effect of merger and radial migration.
Figure~\ref{fig:feagez0merger} decomposes the effect of both. 

The top left panel of Figure~\ref{fig:feagez0merger} shows the AMR for all the stars at redshift z = 0 (thereby we are looking at the combined effects of radial migration + the late merger); the top middle panel shows the stars born before the merger (z = 0.34) at redshift z=0 (thereby we are seeing the effect of radial migration alone, without the merger); and the top right panel shows the stars born before the merger + stars born in the solar neighborhood between z = 0.34 and z = 0 (thereby we are looking at the effect of the merger on the turning point of the AMR). 
The bottom three panels of Figure~\ref{fig:feagez0merger} show the stars in the \alphafe~plane, colored by \dr, for the stars shown in the top panels, respectively. 
Clearly for all selections, the migration is comparable, and there is a remarkable correlation between the radial migration and the [Fe/H]-[$\alpha$/Fe] plane. The top middle panel of Figure~\ref{fig:feagez0merger}, which shows stars that are born before the merger event, suggests that without the satellite infall bringing metal-poor gas, radial migration alone would not be able to create a strong turning point.
However, there is a clear impact of radial migration on the AMR,  after the satellite infall that creates the turning point.
This can be seen by comparing the top left and the top right panels.
Most stars excluded from the right panel in Figure~\ref{fig:feagez0merger} are stars that are born from the metal-poor gas brought in by the merger that are {\it not} born in the current radial bin of 7 $<$ R $<$ 9 kpc.
The stars now at this radius are predominantly born from the diluted gas at other radial bins, and have migrated to the radial bin of interest. 

Since there is a metallicity gradient in the galaxy, migrating stars can widen the turning point along the [Fe/H] direction.
These migrated stars are also born from the gas brought in by the merger, so they are relatively young. 
Therefore, they can also enhance the turning point by migrating to the radial bin of interest, bringing down the average age of stars in the metallicity bins around the turning point. 
However, only 25\% of the stars that are in the turning point have migrated into the radius bin of interest (7 $<$ R $<$ 9 kpc), and 75\% stars in the turning point are created within the current radius bin.

It is also worth pointing out that most old \lowalpha~stars migrated into the the radius bin of interest, as most \lowalpha~stars are born in the galaxy's outskirts (see Figure~\ref{fig:feagez0merger} bottom plots as well as Figure A2 in \cite{Buck2020}), thus showing negative \dr.

We can also understand the connection between the AMR of the stars and the ISM evolution in Figure~\ref{fig:feage_ism}, where we overlay the AMR for stars (black dots) for 7 kpc $<$ R $<$ 9 kpc and $|$z$| <$ 0.5 kpc with the ISM evolution at different current radial bins, taken from Figure.~8 of \cite{Buck2020}. 
The ISM evolution that is closest to the radial bin of the stars is shown in bold (R = 6-8 kpc), and the horizontal lines show the bin edges in metallicity used to calculate the AMR.
The vertical grey dashed line at look-back time$\sim$ 4 Gyr (which is the time in the past from the present-day) shows the time when the satellite made its first passage.
We note two things: While in general the ISM metallicity evolution is monotonically increasing with time (or equivalently with decreasing stellar age), there is a pronounced, rapid decrease in ISM metallicity at the time of the merger.
Note, the merger decreases the metallicity of the ISM at all radii. Since newly born stars at each instance of time inherit the metalllicity of the ISM, all stars younger than $\sim4$ Gyr are born with lower metalllicity than older stars (up to stellar ages of $\sim 8$ Gyr in our simulation where the ISM metalllicity at each radius was as low as after the merger). 
The AMR is derived by calculating the average age in bins of metallicity which is horizontally averaging the age of stars in Fig.~\ref{fig:feage_ism} to create the AMR. 
Thus, those young stars created during the merger moved the apex of the turning point to a lower age by lowering the average age of stars for all metallicity bins that are affected by the merger (e.g. bins that are more metal enriched than $\sim$ -0.1 dex).

In conclusion, both radial migration and late merger events can contribute to the formation of the turnover points.
The detail of how much each mechanism gives rise to the turnover point depends on the initial metallicity gradient in the disk, migration strength, amount of gas brought in from the merger, and possibly other effects.  
We have shown in this section that (i) A late satellite infall during the quiescent phase of galaxy evolution with a mass ratio as low as that for Sagittarius in the MW can bring in enough metal-poor gas to create the turnover point (See Figure~\ref{fig:feagez034});
(ii) the turning points in the AMR are then broadened along the [Fe/H] axis as stars from all over the disk migrate into the radial bin of interest. 
Presumably, in general, the impact of migration versus infall on the AMR turing points depends on the initial metallicity gradient in the disk, the migration strength, the amount and enrichment of the gas from the merger and possibly other perturbing effects on the galaxy. 

\section{Discussion \& future work}\label{sec:dis}
\subsection{Interpreting data with simulation}

With the analysis of a MW-like simulation, we can infer possible pathways to the observed AMR (see Figure 7 of \cite{Lu2021}), in the context of the MW's formation history. 
We highlight that these interpretations that link observational structure to formation mechanisms are based on the assumption that the formation history of the simulation resembles at least some features of the MW's formation history.

First, as discussed in Section~\ref{subsec:feage}, late satellite infall is responsible for creating the AMR turnover points in this simulation.
These turnovers can then be widened along the [Fe/H] axis by radial migration. The stars around the turning points are composed of 1) stars formed from the metal-poor gas brought in by the merger in each R-$|z|$ spatial bin, creating the turnover points, and 2) stars that have formed from the metal-poor gas in other spatial bins, and migrated to the spatial bin of interest.
For stars at the turnover point in the spatial bin with a mean (R, $|z|$) = (8 kpc, 0.25 kpc), at redshift z=0, are comprised of 75\% of stars formed via mechanism 1, and 25\% of stars that have migrated described in mechanism 2.
Since for any given stellar age, the stars that have migrated into the spatial bin are formed at a different mean metallicity (due to the metallicity gradient in the galaxy), they also widen the turning points along the [Fe/H] axis.
This simulation indicates that in the absence of satellite infall, the turning point will not be significant (see the middle panel of Figure~\ref{fig:feagez0merger}).
However, it is not absolutely clear which mechanism contributes more to the turning point in our own Galaxy today. 
This is because the migration strength in the simulation may not be representative of the MW.
However, we measure the overall strength of migration in this simulation and compare it to that measured for the MW in Figure~\ref{fig:migrationhist} in the Appendix. 
We subsequently find a similar shape of the migration between the MW and the simulation overtime, with the simulation showing slightly stronger radial migration across all time until the most recent satellite infall. 
However, it is worth pointing out that the simulation length scale is around two times larger than the MW. 

We now focus on the two major differences between the simulation and the MW data (Figure~7 in \cite{Lu2021}). 1) the turning points exist in both the high- and \lowalpha disk in observations but only the \lowalpha disk in the simulation, and 2) the turnover points have wider openings in the data compared to simulations. 

For 1), The turning points in the \ha\ disk could indicate that radial migration is stronger in the \ha\ disk for the MW than for the \ha\ disk in the simulation.
Satellite infall is still a possibility to create a turnover point in the \ha\ disk in the data. Presumably, however, this will only be possible if any MW merger event and associated gas infall happened before the \lowalpha disk was formed, during the (earlier) \ha~disk formation, and after the epoch when mergers are frequent. 
We note that in the simulation, the gas forming the \ha\ disk stars transitioned entirely to that forming the \lowalpha disk stars (similar to the two-infall model in \cite{Chiappini1997}). Therefore, no new stars can be formed in the \ha\ disk after the \lowalpha disk is formed. 
This may not be the case in the MW. 

For 2), the wider openings in the turnover points in the data along the [Fe/H] axis could also be caused by stronger migration strength in the MW (see Section~\ref{subsec:feage}), uncertainties in the data, and/or the difference in element production in the simulation and the MW.

\begin{figure*}[!htp]
    \centering
    \includegraphics[width=\textwidth]{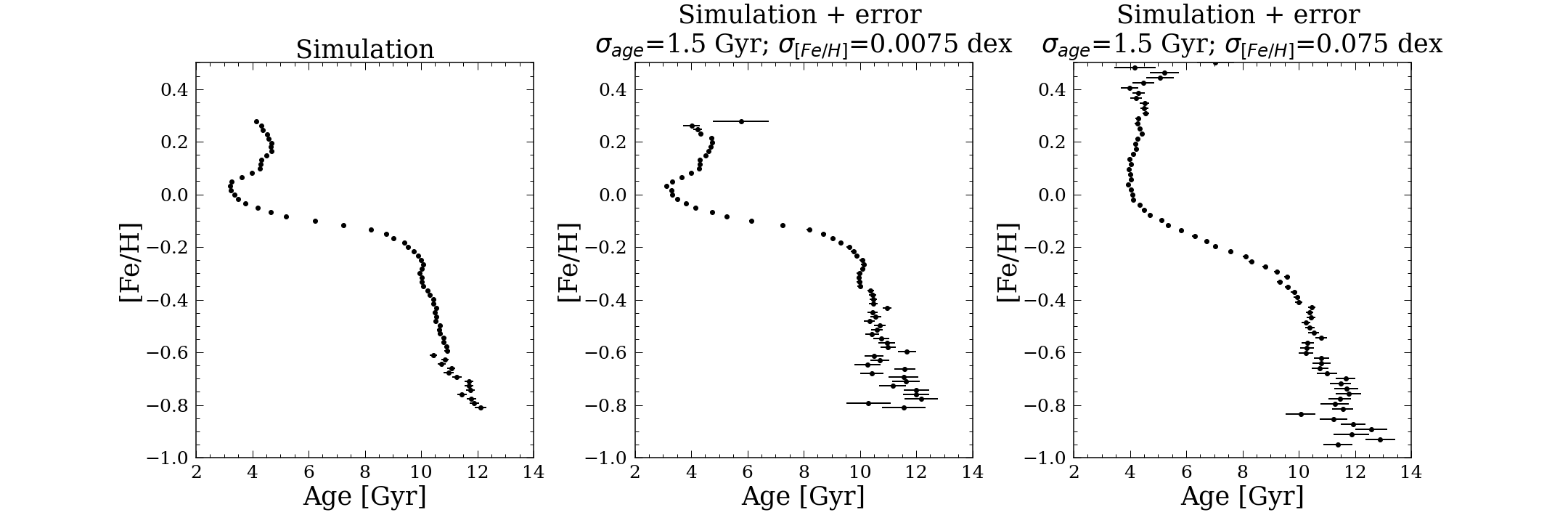}
    \caption{The AMR for stars in the simulation with 7 kpc $<$ R $<$ 9 kpc, and $|z| <$ 0.5 kpc.
    The left plot shows the simulation without perturbing points with the median measurement error, the middle plot shows the same but perturbing each simulation abundance and age with the typical measurement errors (0.0075 dex for [Fe/H], and 2.96 Gyr for stellar age), and the right plot shows the same as the middle plot but increase the uncertainty in [Fe/H] by 10$\times$.
    The errorbars are calculated by taking the standard deviation of the ages in each metallicity bin.}
    \label{fig:feageerr}
\end{figure*}

Figure~\ref{fig:feageerr} shows the effect of measurement uncertainties on the AMR for stars with 7 kpc $<$ R $<$ 9 kpc, and $|z| <$ 0.5 kpc.
We perturbed each metallicity and age in the simulation with a Gaussian with the standard deviation as the median uncertainty of the data 100 times (where the mean uncertainties are $\sigma_{\rm [Fe/H]}$ = 0.0075 dex for [Fe/H] \footnote{The low uncertainty on [Fe/H] is because in \cite{Lu2021}, we only selected stars with metallicity uncertainty $\sigma_{\rm [Fe/H]}$ $<$ 0.03 dex.}, and $\sigma_{\rm age}$ = 1.5 Gyr for stellar age) and calculated the mean [Fe/H] and age to be the new data.
The errorbars in Figure~\ref{fig:feageerr} are calculated by measuring the standard deviation of the ages in each metallicity bin. 
It is clear that the gap does not widen just by perturbing the points with the measurement error.
However, by increasing the metallicity errors by a factor of 10, we are able to produce a wider turnover openings, which is observed in the data. 

As shown in the the right panel in Figure~\ref{fig:feageerr}, with an average uncertainty 10 times more than the actual data (not realistic for APOGEE), the opening of the turning point widens significantly. 
The age uncertainty mostly contributes to the errorbars and not the widening of the turnovers. 
The differences in the widths of the turning points between simulation and data could also be due to the differences between the simulation and MW  metallicity production.

As mentioned in the previous Section (Section \ref{subsec:feage}), the turnover points, created by satellite infall, will be widened by the combination of radial migration and natural enrichment of the star-forming gas. 
If the turning points of the AMR of the MW are created by the same mechanism (satellite infall) as the simulation, a number of additional constraints can be made on the infall in the MW. 
The first interaction of Sagittarius with the MW is thought to have happened in the MW $\sim$ 6 $-$ 8 Gyr ago \citep[e.g.][]{Laporte2018}. 
Therefore, it is likely that the turnover in the MW AMR, if associated with the Sagittarius infall, would indeed be slightly wider than the simulation, where the first interaction of the infall happened around 4 Gyr ago. This is because over time, stars have more time to migrate, and we have seen in the simulation that migration alone in the presence of a turning point widens the turnover. 
As the Sagittarius dwarf is believed to have a similar mass ratio as the satellite galaxy in the simulation, the effect of Sagittarius on the AMR can be expected to be very similar to that in the simulation.
However, the direct comparison between the simulation and Sagittarius is complicated since Sagittarius likely made several passages of the disc over many Gyrs.
Sagittarius was also more massive in the past, so the mass ratio at its first passage was likely higher than the observed mass today. 
In general, the infall orbit of the satellite in the simulation is not a perfect representation of Sagittarius, nor the MW assembly history. 
However, we are not aiming to reproduce the MW AMR precisely and quantatively. Rather, we have sought to understand the origin of the turnover points in the MW and how they change over time. We find that these can result from satellite infall. 

There are also subtle similarities between the simulation and the data (Figure~\ref{fig:feagez0}). 
For example, the possible second turning point in the AMR at the highest metallicity for some radial bins (see also Figure 7 in \cite{Lu2021}).
This turnover point in the AMR of the MW, if real, could be evidence that the turning points observed in the MW are caused by (a number of) satellite infall events, bringing in metal-poor gas and triggering star formation that offsets the existing AMR, as seen in this simulation.

\subsection{Future work}
With future high-resolution spectroscopy surveys such as  MW Mapper \citep{MWM},  we can expect to be able to infer  precise stellar metallicities and ages, for about 10 times more stars than APOGEE \citep{Majewski2017}.
With these data, and sampling that is continuous and contiguous across the plane, we will be able to investigate the very high and low metallicity limit of the current data, across the entire disk and into the bulge.
By comparing the simulations and data, one might be able to distinguish, in detail, the signatures in the observed AMR distributions due to satellite infall versus radial migration.

\appendix \label{sec:append}
\renewcommand{\thesubsection}{\Alph{subsection}}
\counterwithin{figure}{subsection}
\counterwithin{table}{subsection}
\subsection{Additional figures}
    
Figure~\ref{fig:fehz0_hilow} shows the same figure as Figure~\ref{fig:feagez0} but separated by the high and \lowalpha~disks.
It is clear that the major turnover points only exist in the \lowalpha~disk, as expected, since the satellite infall that causes the turnover points happened after the \lowalpha~has formed. 

\begin{figure*}[htp]
    \centering
    \includegraphics[width=\textwidth]{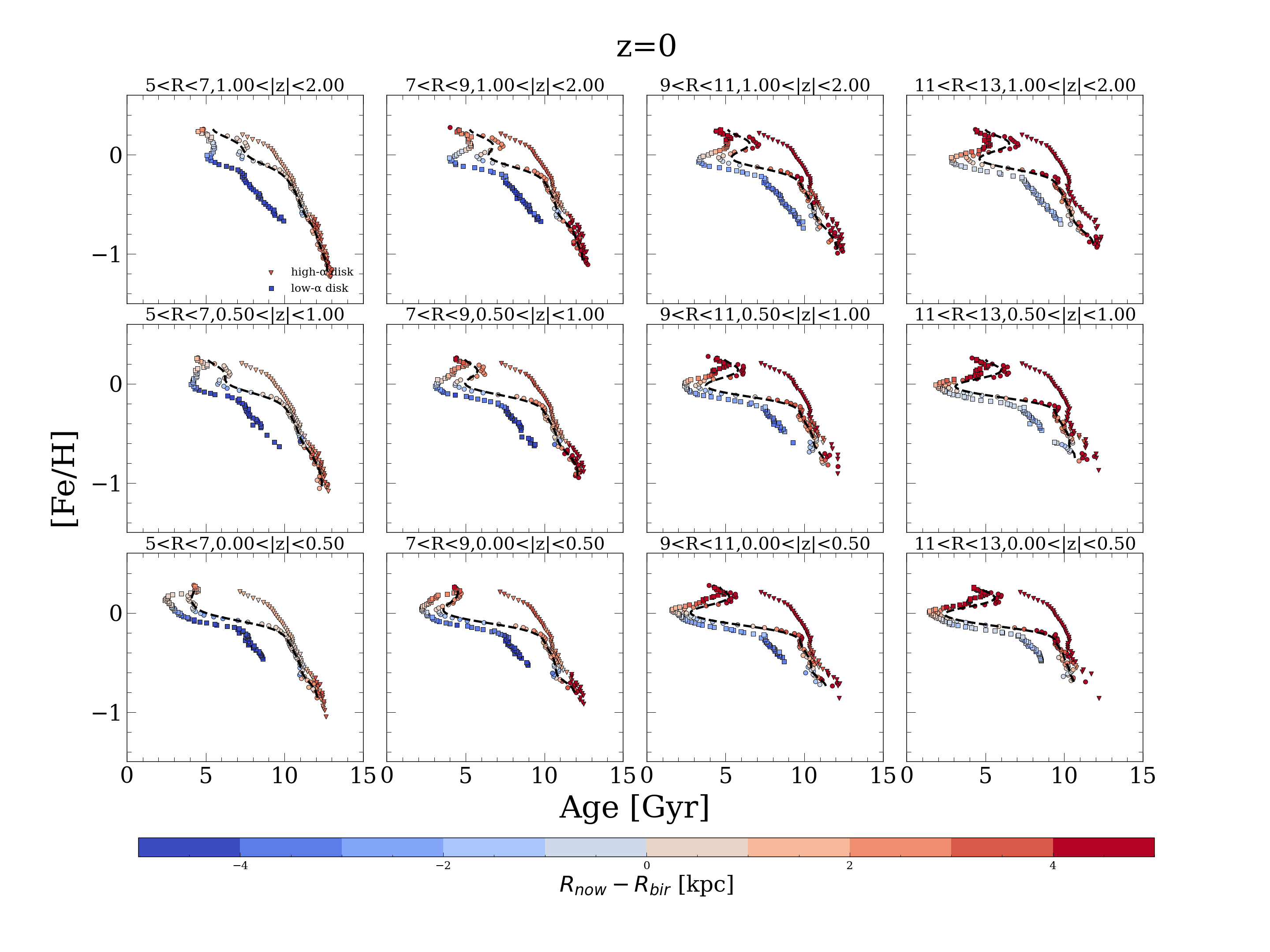}
    \caption{Same as Figure~\ref{fig:feagez0} but separated into the high- and \lowalpha disks. The square markers show the \lowalpha disk stars, the triangles show the \highalpha\ disk stars, and the circles show all stars together. }
    \label{fig:fehz0_hilow}
\end{figure*}

Figure~\ref{fig:fehz0_nomig} shows the stars that are born and remain within each spatial bin of Figure~\ref{fig:feagez0}, thereby illustrating the AMR in the absence of migration. Note the AMR turning points are all still present, as the migration does not create these. 

\begin{figure*}[htp]
    \centering
    \includegraphics[width=\textwidth]{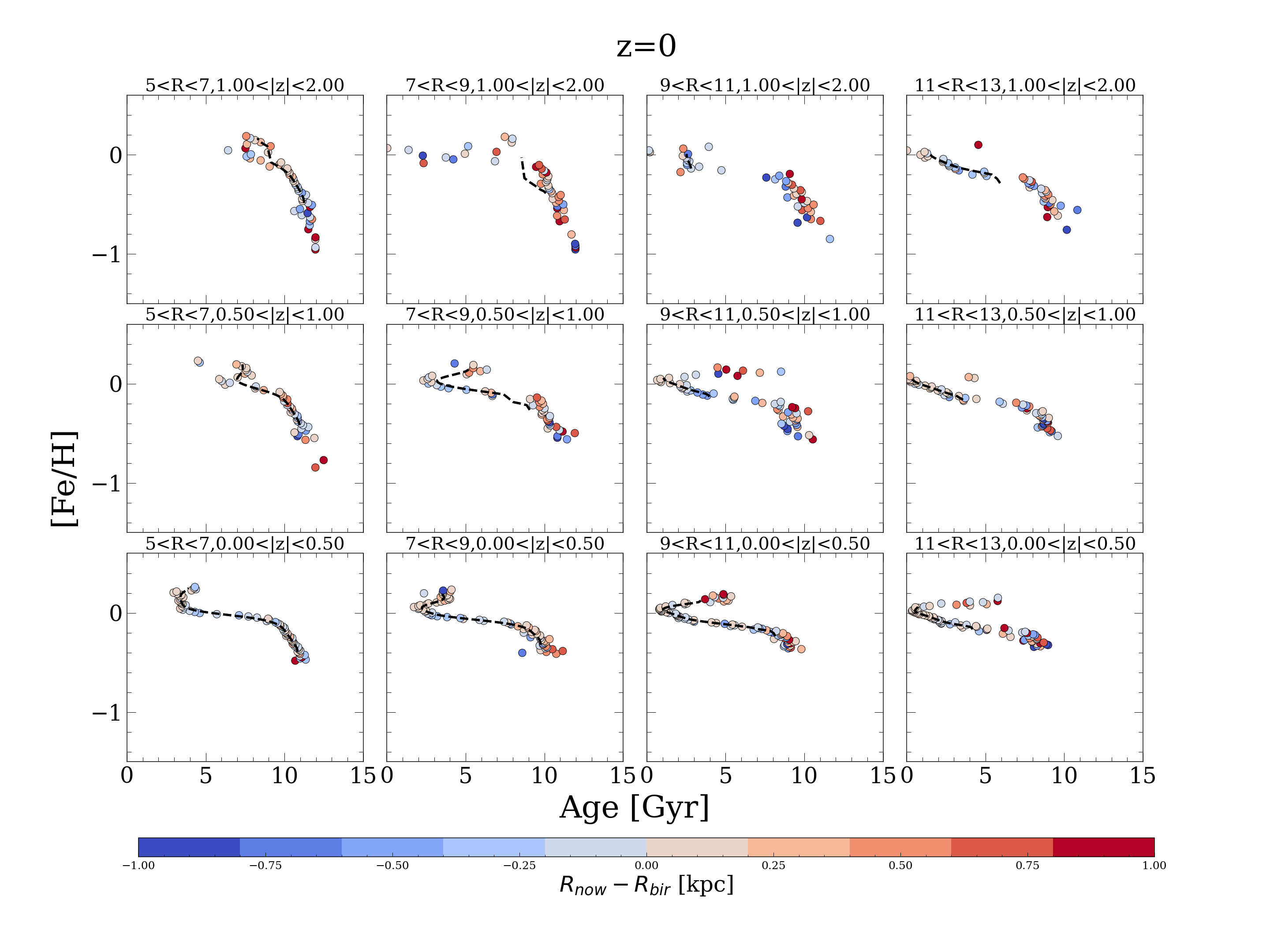}
    \caption{Same as Figure~\ref{fig:feagez0} but only for stars that are born within each spatial bin. 
    It is clear that without radial migration, the turnover points still exist in most bins. 
    However, the most metal-enriched and metal-poor part of the current day AMR (see Figure~\ref{fig:feagez0}) migrated from other spatial bins as expected. }
    \label{fig:fehz0_nomig}
\end{figure*}

Figure~\ref{fig:fehsnap} shows the metallicity distribution of stars born between each snapshot around the merger in the radial bin of R = 7 $-$ 9 kpc with galactic height $|z| <$ 0.5 kpc. 
For each snapshot, we calculated the mean (points) and standard deviation (errorbars) in [Fe/H] for the newly born stars.
The shaded area in gray marks the approximate time of the first passage of the infall (where Age=0 is the present day at redshift 0).
It is clear that when the merger happened, stars started to form at a lower metallicity bin as the infall brought in metal-poor gas (also shown in the dip in ISM metallicity in Figure~\ref{fig:feage_ism} around 4 Gyr), causing the turnover point.

\begin{figure*}[htp]
    \centering
    \includegraphics[width=\textwidth]{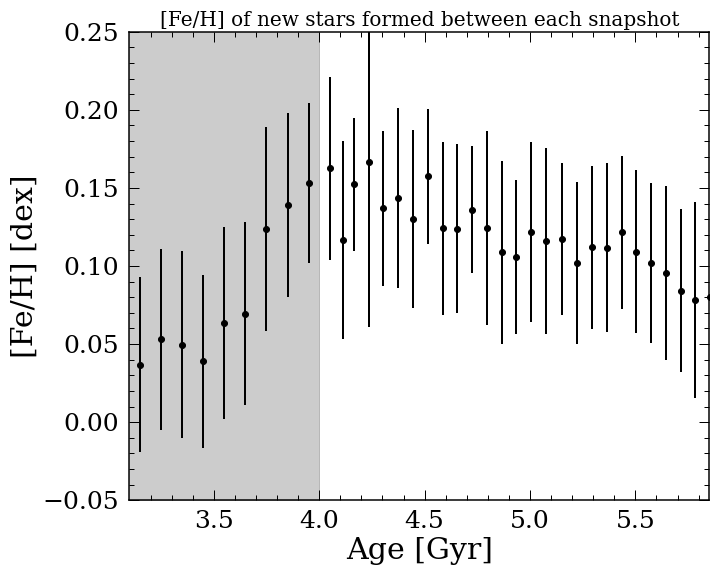}
    \caption{The mean (points) and standard deviation (errorbars) of metallicity of stars born between each snapshot ($\sim$ 200 Myr apart) around the satellite infall in the radial bin of 7-9 kpc with galactic height $|z| <$ 0.5 kpc.
    The gray area marked the approximate time of the satellite infall (first passage around 4 Gyr). 
    Combined with Figure~\ref{fig:feage_ism}, we can see the decrease in mean metallicity of stars born after the merger, creating the turnover point.}
    \label{fig:fehsnap}
\end{figure*}

Figure~\ref{fig:migrationhist} shows the migration strength in different age bin for the stars in the simulation (black dots) and the model in \cite{Frankel2019} (3.9 kpc $(\tau/7 Gyr)^0.5$; black line) for disk stars (selected by choosing only stars with radius $>$ 5 kpc). 
We calculated the strength ($\sigma$) by taking the standard deviation of \dr~in each age bin for $R > $ 5 kpc to exclude the bulge stars.
The shape of the migration strength in the simulation matches with that from \cite{Frankel2019} until the most recent merger with a relatively constant offset.

\begin{figure*}[htp]
    \centering
    \includegraphics[width=0.5\textwidth]{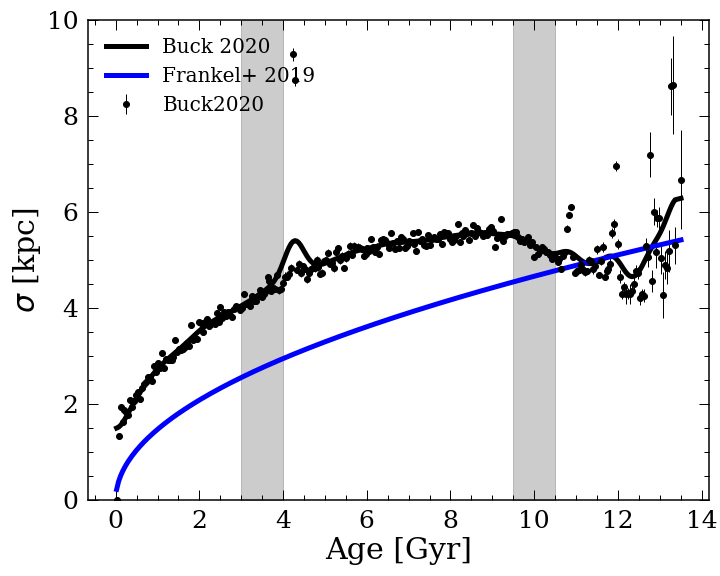}
    \caption{The radial migration strength for disk stars as function of age.
    The black dots show the radial migration strength in different age bin for the simulation (with $R >$ 5 kpc), the black line shows the smoothed curve over the black dots, and the blue line shows the migration strength for the MW as inferred in Frankel et al. (2019) (3.9 kpc $(\tau/7 Gyr)^0.5$). The grey shaded area shows the approximate time of the two most recent merger events ($\sim$ 3 and 10 Gyrs ago). }
    \label{fig:migrationhist}
\end{figure*}

\section*{Acknowledgements}
Melissa Ness is supported in part by a Sloan Fellowship. 

Tobias Buck acknowledges support from the European Research Council under ERC-CoG grant CRAGSMAN-646955. This research made use of {\sc{pynbody}} \citet{pynbody}.
We gratefully acknowledge the Gauss Centre for Supercomputing e.V. (www.gauss-centre.eu) for funding this project by providing computing time on the GCS Supercomputer SuperMUC at Leibniz Supercomputing Centre (www.lrz.de).
This research was carried out on the High Performance Computing resources at New York University Abu Dhabi. We greatly appreciate the contributions of all these computing allocations.

This paper includes data collected by the {\it Kepler} mission. Funding for the {\it Kepler} mission is provided by the NASA Science Mission directorate.

This work has made use of data from the European Space Agency (ESA) mission
{\it Gaia} (\url{https://www.cosmos.esa.int/gaia}), processed by the {\it Gaia}
Data Processing and Analysis Consortium (DPAC,
\url{https://www.cosmos.esa.int/web/gaia/dpac/consortium}). Funding for the DPAC
has been provided by national institutions, in particular the institutions
participating in the {\it Gaia} Multilateral Agreement.

Funding for the Sloan Digital Sky 
Survey IV has been provided by the 
Alfred P. Sloan Foundation, the U.S. 
Department of Energy Office of 
Science, and the Participating 
Institutions. 

SDSS-IV acknowledges support and 
resources from the Center for High 
Performance Computing  at the 
University of Utah. The SDSS 
website is www.sdss.org.

SDSS-IV is managed by the 
Astrophysical Research Consortium 
for the Participating Institutions 
of the SDSS Collaboration including 
the Brazilian Participation Group, 
the Carnegie Institution for Science, 
Carnegie Mellon University, Center for 
Astrophysics | Harvard \& 
Smithsonian, the Chilean Participation 
Group, the French Participation Group, 
Instituto de Astrof\'isica de 
Canarias, The Johns Hopkins 
University, Kavli Institute for the 
Physics and Mathematics of the 
Universe (IPMU) / University of 
Tokyo, the Korean Participation Group, 
Lawrence Berkeley National Laboratory, 
Leibniz Institut f\"ur Astrophysik 
Potsdam (AIP),  Max-Planck-Institut 
f\"ur Astronomie (MPIA Heidelberg), 
Max-Planck-Institut f\"ur 
Astrophysik (MPA Garching), 
Max-Planck-Institut f\"ur 
Extraterrestrische Physik (MPE), 
National Astronomical Observatories of 
China, New Mexico State University, 
New York University, University of 
Notre Dame, Observat\'ario 
Nacional / MCTI, The Ohio State 
University, Pennsylvania State 
University, Shanghai 
Astronomical Observatory, United 
Kingdom Participation Group, 
Universidad Nacional Aut\'onoma 
de M\'exico, University of Arizona, 
University of Colorado Boulder, 
University of Oxford, University of 
Portsmouth, University of Utah, 
University of Virginia, University 
of Washington, University of 
Wisconsin, Vanderbilt University, and Yale University.

Guoshoujing Telescope (the Large Sky Area Multi-Object Fiber Spectroscopic Telescope LAMOST) is a National Major Scientific Project built by the Chinese Academy of Sciences. Funding for the project has been provided by the National Development and Reform Commission. LAMOST is operated and managed by the National Astronomical Observatories, Chinese Academy of Sciences.

This research made use of Astropy,\footnote{http://www.astropy.org} a community-developed core Python package for Astronomy \citep{astropy:2013, astropy:2018}.

\vspace{5mm}
Facilities: Gaia, Kepler, APOGEE


Softwares: Astropy \citep{astropy:2013, astropy:2018}, Numpy \citep{oliphant2006guide}, sklearn \citep{scikit-learn}, The Cannon \citep{Ness2015}, Astraea \citep{Lu2020}.



\bibliographystyle{mnras}
\bibliography{references.bib} 

\begin{thebibliography}{}
\makeatletter
\relax
\def\mn@urlcharsother{\let\do\@makeother \do\$\do\&\do\#\do\^\do\_\do\%\do\~}
\def\mn@doi{\begingroup\mn@urlcharsother \@ifnextchar [ {\mn@doi@}
  {\mn@doi@[]}}
\def\mn@doi@[#1]#2{\def\@tempa{#1}\ifx\@tempa\@empty \href
  {http://dx.doi.org/#2} {doi:#2}\else \href {http://dx.doi.org/#2} {#1}\fi
  \endgroup}
\def\mn@eprint#1#2{\mn@eprint@#1:#2::\@nil}
\def\mn@eprint@arXiv#1{\href {http://arxiv.org/abs/#1} {{\tt arXiv:#1}}}
\def\mn@eprint@dblp#1{\href {http://dblp.uni-trier.de/rec/bibtex/#1.xml}
  {dblp:#1}}
\def\mn@eprint@#1:#2:#3:#4\@nil{\def\@tempa {#1}\def\@tempb {#2}\def\@tempc
  {#3}\ifx \@tempc \@empty \let \@tempc \@tempb \let \@tempb \@tempa \fi \ifx
  \@tempb \@empty \def\@tempb {arXiv}\fi \@ifundefined
  {mn@eprint@\@tempb}{\@tempb:\@tempc}{\expandafter \expandafter \csname
  mn@eprint@\@tempb\endcsname \expandafter{\@tempc}}}

\bibitem[\protect\citeauthoryear{{Astropy Collaboration} et~al.,}{{Astropy
  Collaboration} et~al.}{2013}]{astropy:2013}
{Astropy Collaboration} et~al., 2013, \mn@doi [\aap]
  {10.1051/0004-6361/201322068}, \href
  {http://adsabs.harvard.edu/abs/2013A%26A...558A..33A} {558, A33}

\bibitem[\protect\citeauthoryear{{Aumer}, {White}, {Naab}  \&
  {Scannapieco}}{{Aumer} et~al.}{2013}]{Aumer2013}
{Aumer} M.,  {White} S. D.~M.,  {Naab} T.,   {Scannapieco} C.,  2013, \mn@doi
  [\mnras] {10.1093/mnras/stt1230}, \href
  {https://ui.adsabs.harvard.edu/abs/2013MNRAS.434.3142A} {434, 3142}

\bibitem[\protect\citeauthoryear{{Bird}, {Kazantzidis}  \& {Weinberg}}{{Bird}
  et~al.}{2012}]{Bird2012}
{Bird} J.~C.,  {Kazantzidis} S.,   {Weinberg} D.~H.,  2012, \mn@doi [\mnras]
  {10.1111/j.1365-2966.2011.19728.x}, \href
  {https://ui.adsabs.harvard.edu/abs/2012MNRAS.420..913B} {420, 913}

\bibitem[\protect\citeauthoryear{{Bland-Hawthorn} \&
  {Gerhard}}{{Bland-Hawthorn} \& {Gerhard}}{2016}]{Bland2016}
{Bland-Hawthorn} J.,  {Gerhard} O.,  2016, \mn@doi [\araa]
  {10.1146/annurev-astro-081915-023441}, \href
  {https://ui.adsabs.harvard.edu/abs/2016ARA&A..54..529B} {54, 529}

\bibitem[\protect\citeauthoryear{{Bovy}, {Rix}, {Liu}, {Hogg}, {Beers}  \&
  {Lee}}{{Bovy} et~al.}{2012}]{Bovy2012}
{Bovy} J.,  {Rix} H.-W.,  {Liu} C.,  {Hogg} D.~W.,  {Beers} T.~C.,   {Lee}
  Y.~S.,  2012, \mn@doi [\apj] {10.1088/0004-637X/753/2/148}, \href
  {https://ui.adsabs.harvard.edu/abs/2012ApJ...753..148B} {753, 148}

\bibitem[\protect\citeauthoryear{Buck}{Buck}{2019}]{Buck2020}
Buck T.,  2019, \mn@doi [Monthly Notices of the Royal Astronomical Society]
  {10.1093/mnras/stz3289}, 491, 5435

\bibitem[\protect\citeauthoryear{{Buck}, {Ness}, {Macci{\`o}}, {Obreja}  \&
  {Dutton}}{{Buck} et~al.}{2018}]{Buck2018}
{Buck} T.,  {Ness} M.~K.,  {Macci{\`o}} A.~V.,  {Obreja} A.,   {Dutton} A.~A.,
  2018, \mn@doi [\apj] {10.3847/1538-4357/aac890}, \href
  {https://ui.adsabs.harvard.edu/abs/2018ApJ...861...88B} {861, 88}

\bibitem[\protect\citeauthoryear{{Buck}, {Macci{\`o}}, {Dutton}, {Obreja}  \&
  {Frings}}{{Buck} et~al.}{2019a}]{Buck2019b}
{Buck} T.,  {Macci{\`o}} A.~V.,  {Dutton} A.~A.,  {Obreja} A.,   {Frings} J.,
  2019a, \mn@doi [\mnras] {10.1093/mnras/sty2913}, \href
  {https://ui.adsabs.harvard.edu/abs/2019MNRAS.483.1314B} {483, 1314}

\bibitem[\protect\citeauthoryear{{Buck}, {Dutton}  \& {Macci{\`o}}}{{Buck}
  et~al.}{2019b}]{Buck2019c}
{Buck} T.,  {Dutton} A.~A.,   {Macci{\`o}} A.~V.,  2019b, \mn@doi [\mnras]
  {10.1093/mnras/stz969}, \href
  {https://ui.adsabs.harvard.edu/abs/2019MNRAS.486.1481B} {486, 1481}

\bibitem[\protect\citeauthoryear{{Buck}, {Ness}, {Obreja}, {Macci{\`o}}  \&
  {Dutton}}{{Buck} et~al.}{2019c}]{Buck2019}
{Buck} T.,  {Ness} M.,  {Obreja} A.,  {Macci{\`o}} A.~V.,   {Dutton} A.~A.,
  2019c, \mn@doi [\apj] {10.3847/1538-4357/aaffd0}, \href
  {https://ui.adsabs.harvard.edu/abs/2019ApJ...874...67B} {874, 67}

\bibitem[\protect\citeauthoryear{{Buck}, {Obreja}, {Macci{\`o}}, {Minchev},
  {Dutton}  \& {Ostriker}}{{Buck} et~al.}{2020}]{Buck2020b}
{Buck} T.,  {Obreja} A.,  {Macci{\`o}} A.~V.,  {Minchev} I.,  {Dutton} A.~A.,
  {Ostriker} J.~P.,  2020, \mn@doi [\mnras] {10.1093/mnras/stz3241}, \href
  {https://ui.adsabs.harvard.edu/abs/2020MNRAS.491.3461B} {491, 3461}

\bibitem[\protect\citeauthoryear{{Buder} et~al.,}{{Buder}
  et~al.}{2019}]{Buder2019}
{Buder} S.,  et~al., 2019, \mn@doi [\aap] {10.1051/0004-6361/201833218}, \href
  {https://ui.adsabs.harvard.edu/abs/2019A&A...624A..19B} {624, A19}

\bibitem[\protect\citeauthoryear{{Chiappini}, {Matteucci}  \&
  {Gratton}}{{Chiappini} et~al.}{1997}]{Chiappini1997}
{Chiappini} C.,  {Matteucci} F.,   {Gratton} R.,  1997, \mn@doi [\apj]
  {10.1086/303726}, \href
  {https://ui.adsabs.harvard.edu/abs/1997ApJ...477..765C} {477, 765}

\bibitem[\protect\citeauthoryear{{Chiappini} et~al.,}{{Chiappini}
  et~al.}{2015}]{Chiappini2015}
{Chiappini} C.,  et~al., 2015, \mn@doi [\aap] {10.1051/0004-6361/201525865},
  \href {https://ui.adsabs.harvard.edu/abs/2015A&A...576L..12C} {576, L12}

\bibitem[\protect\citeauthoryear{{Clarke} et~al.,}{{Clarke}
  et~al.}{2019}]{Clarke2019}
{Clarke} A.~J.,  et~al., 2019, \mn@doi [\mnras] {10.1093/mnras/stz104}, \href
  {https://ui.adsabs.harvard.edu/abs/2019MNRAS.484.3476C} {484, 3476}

\bibitem[\protect\citeauthoryear{{Cui} et~al.,}{{Cui} et~al.}{2012}]{LAMOST}
{Cui} X.-Q.,  et~al., 2012, \mn@doi [Research in Astronomy and Astrophysics]
  {10.1088/1674-4527/12/9/003}, \href
  {https://ui.adsabs.harvard.edu/abs/2012RAA....12.1197C} {12, 1197}

\bibitem[\protect\citeauthoryear{{Daniel} \& {Wyse}}{{Daniel} \&
  {Wyse}}{2018}]{Daniel2018}
{Daniel} K.~J.,  {Wyse} R. F.~G.,  2018, \mn@doi [\mnras]
  {10.1093/mnras/sty199}, \href
  {https://ui.adsabs.harvard.edu/abs/2018MNRAS.476.1561D} {476, 1561}

\bibitem[\protect\citeauthoryear{{De Silva} et~al.,}{{De Silva}
  et~al.}{2015}]{Silva2015}
{De Silva} G.~M.,  et~al., 2015, \mn@doi [\mnras] {10.1093/mnras/stv327}, \href
  {https://ui.adsabs.harvard.edu/abs/2015MNRAS.449.2604D} {449, 2604}

\bibitem[\protect\citeauthoryear{{Dutton}, {Macci{\`o}}, {Buck}, {Dixon},
  {Blank}  \& {Obreja}}{{Dutton} et~al.}{2019}]{Dutton2019}
{Dutton} A.~A.,  {Macci{\`o}} A.~V.,  {Buck} T.,  {Dixon} K.~L.,  {Blank} M.,
  {Obreja} A.,  2019, \mn@doi [\mnras] {10.1093/mnras/stz889}, \href
  {https://ui.adsabs.harvard.edu/abs/2019MNRAS.486..655D} {486, 655}

\bibitem[\protect\citeauthoryear{{Feuillet} et~al.,}{{Feuillet}
  et~al.}{2018}]{Feuillet2018}
{Feuillet} D.~K.,  et~al., 2018, \mn@doi [\mnras] {10.1093/mnras/sty779}, \href
  {https://ui.adsabs.harvard.edu/abs/2018MNRAS.477.2326F} {477, 2326}

\bibitem[\protect\citeauthoryear{{Feuillet}, {Frankel}, {Lind}, {Frinchaboy},
  {Garc{\'\i}a-Hern{\'a}ndez}, {Lane}, {Nitschelm}  \&
  {Roman-Lopes}}{{Feuillet} et~al.}{2019}]{Feuillet2019}
{Feuillet} D.~K.,  {Frankel} N.,  {Lind} K.,  {Frinchaboy} P.~M.,
  {Garc{\'\i}a-Hern{\'a}ndez} D.~A.,  {Lane} R.~R.,  {Nitschelm} C.,
  {Roman-Lopes} A.,  2019, \mn@doi [\mnras] {10.1093/mnras/stz2221}, \href
  {https://ui.adsabs.harvard.edu/abs/2019MNRAS.489.1742F} {489, 1742}

\bibitem[\protect\citeauthoryear{{Frankel}, {Sanders}, {Rix}, {Ting}  \&
  {Ness}}{{Frankel} et~al.}{2019}]{Frankel2019}
{Frankel} N.,  {Sanders} J.,  {Rix} H.-W.,  {Ting} Y.-S.,   {Ness} M.,  2019,
  \mn@doi [\apj] {10.3847/1538-4357/ab4254}, \href
  {https://ui.adsabs.harvard.edu/abs/2019ApJ...884...99F} {884, 99}

\bibitem[\protect\citeauthoryear{{Frankel}, {Sanders}, {Ting}  \&
  {Rix}}{{Frankel} et~al.}{2020}]{Frankel2020}
{Frankel} N.,  {Sanders} J.,  {Ting} Y.-S.,   {Rix} H.-W.,  2020, \mn@doi
  [\apj] {10.3847/1538-4357/ab910c}, \href
  {https://ui.adsabs.harvard.edu/abs/2020ApJ...896...15F} {896, 15}

\bibitem[\protect\citeauthoryear{{Gilmore} et~al.,}{{Gilmore}
  et~al.}{2012}]{Gilmore2012}
{Gilmore} G.,  et~al., 2012, The Messenger, \href
  {https://ui.adsabs.harvard.edu/abs/2012Msngr.147...25G} {147, 25}

\bibitem[\protect\citeauthoryear{{Grand} et~al.,}{{Grand}
  et~al.}{2017}]{Grand2017}
{Grand} R. J.~J.,  et~al., 2017, \mn@doi [\mnras] {10.1093/mnras/stx071}, \href
  {https://ui.adsabs.harvard.edu/abs/2017MNRAS.467..179G} {467, 179}

\bibitem[\protect\citeauthoryear{{Grenon}}{{Grenon}}{1972}]{Grenon1972}
{Grenon} M.,  1972, in {Cayrel de Strobel} G.,  {Delplace} A.~M.,  eds, IAU
  Colloq. 17: Age des Etoiles. p.~55

\bibitem[\protect\citeauthoryear{{Grenon}}{{Grenon}}{1989}]{Grenon1989}
{Grenon} M.,  1989, \mn@doi [\apss] {10.1007/BF00646341}, \href
  {https://ui.adsabs.harvard.edu/abs/1989Ap&SS.156...29G} {156, 29}

\bibitem[\protect\citeauthoryear{{Hayden} et~al.,}{{Hayden}
  et~al.}{2015}]{Hayden2015}
{Hayden} M.~R.,  et~al., 2015, \mn@doi [\apj] {10.1088/0004-637X/808/2/132},
  \href {https://ui.adsabs.harvard.edu/abs/2015ApJ...808..132H} {808, 132}

\bibitem[\protect\citeauthoryear{{Hekker} \& {Johnson}}{{Hekker} \&
  {Johnson}}{2019}]{Hekker2019}
{Hekker} S.,  {Johnson} J.~A.,  2019, \mn@doi [\mnras] {10.1093/mnras/stz1554},
  \href {https://ui.adsabs.harvard.edu/abs/2019MNRAS.487.4343H} {487, 4343}

\bibitem[\protect\citeauthoryear{{Hilmi} et~al.,}{{Hilmi}
  et~al.}{2020}]{Hilmi2020}
{Hilmi} T.,  et~al., 2020, \mn@doi [\mnras] {10.1093/mnras/staa1934}, \href
  {https://ui.adsabs.harvard.edu/abs/2020MNRAS.497..933H} {497, 933}

\bibitem[\protect\citeauthoryear{{Hopkins} et~al.,}{{Hopkins}
  et~al.}{2018}]{Hopkins2018}
{Hopkins} P.~F.,  et~al., 2018, \mn@doi [\mnras] {10.1093/mnras/sty1690}, \href
  {https://ui.adsabs.harvard.edu/abs/2018MNRAS.480..800H} {480, 800}

\bibitem[\protect\citeauthoryear{{Ibata}, {Gilmore}  \& {Irwin}}{{Ibata}
  et~al.}{1994}]{Ibata1994}
{Ibata} R.~A.,  {Gilmore} G.,   {Irwin} M.~J.,  1994, \mn@doi [\nat]
  {10.1038/370194a0}, \href
  {https://ui.adsabs.harvard.edu/abs/1994Natur.370..194I} {370, 194}

\bibitem[\protect\citeauthoryear{{Jofr{\'e}} et~al.,}{{Jofr{\'e}}
  et~al.}{2016}]{Jofre2016}
{Jofr{\'e}} P.,  et~al., 2016, \mn@doi [\aap] {10.1051/0004-6361/201629356},
  \href {https://ui.adsabs.harvard.edu/abs/2016A&A...595A..60J} {595, A60}

\bibitem[\protect\citeauthoryear{{Keller}, {Wadsley}, {Wang}  \&
  {Kruijssen}}{{Keller} et~al.}{2019}]{Keller2019}
{Keller} B.~W.,  {Wadsley} J.~W.,  {Wang} L.,   {Kruijssen} J.~M.~D.,  2019,
  \mn@doi [\mnras] {10.1093/mnras/sty2859}, \href
  {https://ui.adsabs.harvard.edu/abs/2019MNRAS.482.2244K} {482, 2244}

\bibitem[\protect\citeauthoryear{{Kollmeier} et~al.,}{{Kollmeier}
  et~al.}{2017}]{MWM}
{Kollmeier} J.~A.,  et~al., 2017, arXiv e-prints, \href
  {https://ui.adsabs.harvard.edu/abs/2017arXiv171103234K} {p. arXiv:1711.03234}

\bibitem[\protect\citeauthoryear{{Kordopatis} et~al.,}{{Kordopatis}
  et~al.}{2015}]{Kordopatis2015}
{Kordopatis} G.,  et~al., 2015, \mn@doi [\mnras] {10.1093/mnras/stu2726}, \href
  {https://ui.adsabs.harvard.edu/abs/2015MNRAS.447.3526K} {447, 3526}

\bibitem[\protect\citeauthoryear{{Laporte}, {Johnston}, {G{\'o}mez},
  {Garavito-Camargo}  \& {Besla}}{{Laporte} et~al.}{2018}]{Laporte2018}
{Laporte} C. F.~P.,  {Johnston} K.~V.,  {G{\'o}mez} F.~A.,  {Garavito-Camargo}
  N.,   {Besla} G.,  2018, \mn@doi [\mnras] {10.1093/mnras/sty1574}, \href
  {https://ui.adsabs.harvard.edu/abs/2018MNRAS.481..286L} {481, 286}

\bibitem[\protect\citeauthoryear{{Larson}}{{Larson}}{1972}]{Larson1972}
{Larson} R.~B.,  1972, \mn@doi [Nature Physical Science]
  {10.1038/physci236007a0}, \href
  {https://ui.adsabs.harvard.edu/abs/1972NPhS..236....7L} {236, 7}

\bibitem[\protect\citeauthoryear{{Loebman}, {Debattista}, {Nidever}, {Hayden},
  {Holtzman}, {Clarke}, {Ro{\v{s}}kar}  \& {Valluri}}{{Loebman}
  et~al.}{2016}]{Loebman2016}
{Loebman} S.~R.,  {Debattista} V.~P.,  {Nidever} D.~L.,  {Hayden} M.~R.,
  {Holtzman} J.~A.,  {Clarke} A.~J.,  {Ro{\v{s}}kar} R.,   {Valluri} M.,  2016,
  \mn@doi [\apjl] {10.3847/2041-8205/818/1/L6}, \href
  {https://ui.adsabs.harvard.edu/abs/2016ApJ...818L...6L} {818, L6}

\bibitem[\protect\citeauthoryear{{Lu}, {Angus}, {Ag{\"u}eros}, {Blancato},
  {Ness}, {Rowland}, {Curtis}  \& {Grunblatt}}{{Lu} et~al.}{2020}]{Lu2020}
{Lu} Y.~L.,  {Angus} R.,  {Ag{\"u}eros} M.~A.,  {Blancato} K.,  {Ness} M.,
  {Rowland} D.,  {Curtis} J.~L.,   {Grunblatt} S.,  2020, \mn@doi [\aj]
  {10.3847/1538-3881/abada4}, \href
  {https://ui.adsabs.harvard.edu/abs/2020AJ....160..168L} {160, 168}

\bibitem[\protect\citeauthoryear{{Lu}, {Ness}, {Buck}  \& {Zinn}}{{Lu}
  et~al.}{2021}]{Lu2021}
{Lu} Y.,  {Ness} M.,  {Buck} T.,   {Zinn} J.,  2021, arXiv e-prints, \href
  {https://ui.adsabs.harvard.edu/abs/2021arXiv210212003Y} {p. arXiv:2102.12003}

\bibitem[\protect\citeauthoryear{{Majewski} et~al.,}{{Majewski}
  et~al.}{2017}]{Majewski2017}
{Majewski} S.~R.,  et~al., 2017, \mn@doi [\aj] {10.3847/1538-3881/aa784d},
  \href {https://ui.adsabs.harvard.edu/abs/2017AJ....154...94M} {154, 94}

\bibitem[\protect\citeauthoryear{{Marinacci}, {Pakmor}  \&
  {Springel}}{{Marinacci} et~al.}{2014}]{Marinacci2014}
{Marinacci} F.,  {Pakmor} R.,   {Springel} V.,  2014, \mn@doi [\mnras]
  {10.1093/mnras/stt2003}, \href
  {https://ui.adsabs.harvard.edu/abs/2014MNRAS.437.1750M} {437, 1750}

\bibitem[\protect\citeauthoryear{{Martig} et~al.,}{{Martig}
  et~al.}{2015}]{Martig2015}
{Martig} M.,  et~al., 2015, \mn@doi [\mnras] {10.1093/mnras/stv1071}, \href
  {https://ui.adsabs.harvard.edu/abs/2015MNRAS.451.2230M} {451, 2230}

\bibitem[\protect\citeauthoryear{{Minchev} et~al.,}{{Minchev}
  et~al.}{2018}]{Minchev2018}
{Minchev} I.,  et~al., 2018, \mn@doi [\mnras] {10.1093/mnras/sty2033}, \href
  {https://ui.adsabs.harvard.edu/abs/2018MNRAS.481.1645M} {481, 1645}

\bibitem[\protect\citeauthoryear{{Munshi}, {Brooks}, {Christensen},
  {Applebaum}, {Holley-Bockelmann}, {Quinn}  \& {Wadsley}}{{Munshi}
  et~al.}{2019}]{Munshi2019}
{Munshi} F.,  {Brooks} A.~M.,  {Christensen} C.,  {Applebaum} E.,
  {Holley-Bockelmann} K.,  {Quinn} T.~R.,   {Wadsley} J.,  2019, \mn@doi [\apj]
  {10.3847/1538-4357/ab0085}, \href
  {https://ui.adsabs.harvard.edu/abs/2019ApJ...874...40M} {874, 40}

\bibitem[\protect\citeauthoryear{{Ness}, {Hogg}, {Rix}, {Ho}  \&
  {Zasowski}}{{Ness} et~al.}{2015}]{Ness2015}
{Ness} M.,  {Hogg} D.~W.,  {Rix} H.~W.,  {Ho} A. Y.~Q.,   {Zasowski} G.,  2015,
  \mn@doi [\apj] {10.1088/0004-637X/808/1/16}, \href
  {https://ui.adsabs.harvard.edu/abs/2015ApJ...808...16N} {808, 16}

\bibitem[\protect\citeauthoryear{Oliphant}{Oliphant}{2006}]{oliphant2006guide}
Oliphant T.~E.,  2006, A guide to NumPy.
~1 Vol. 1, Trelgol Publishing USA

\bibitem[\protect\citeauthoryear{Pedregosa et~al.,}{Pedregosa
  et~al.}{2011}]{scikit-learn}
Pedregosa F.,  et~al., 2011, Journal of Machine Learning Research, 12, 2825

\bibitem[\protect\citeauthoryear{{Pontzen}, {Ro{\v s}kar}, {Stinson}, {Woods},
  {Reed}, {Coles}  \& {Quinn}}{{Pontzen} et~al.}{2013}]{pynbody}
{Pontzen} A.,  {Ro{\v s}kar} R.,  {Stinson} G.~S.,  {Woods} R.,  {Reed} D.~M.,
  {Coles} J.,   {Quinn} T.~R.,  2013, {pynbody: Astrophysics Simulation
  Analysis for Python}

\bibitem[\protect\citeauthoryear{{Price-Whelan} et~al.,}{{Price-Whelan}
  et~al.}{2018}]{astropy:2018}
{Price-Whelan} A.~M.,  et~al., 2018, \mn@doi [\aj] {10.3847/1538-3881/aabc4f},
  \href {https://ui.adsabs.harvard.edu/#abs/2018AJ....156..123T} {156, 123}

\bibitem[\protect\citeauthoryear{{Queiroz} et~al.,}{{Queiroz}
  et~al.}{2018}]{Queiroz2018}
{Queiroz} A.~B.~A.,  et~al., 2018, \mn@doi [\mnras] {10.1093/mnras/sty330},
  \href {https://ui.adsabs.harvard.edu/abs/2018MNRAS.476.2556Q} {476, 2556}

\bibitem[\protect\citeauthoryear{{Quillen}, {Minchev}, {Bland-Hawthorn}  \&
  {Haywood}}{{Quillen} et~al.}{2009}]{Quillen2009}
{Quillen} A.~C.,  {Minchev} I.,  {Bland-Hawthorn} J.,   {Haywood} M.,  2009,
  \mn@doi [\mnras] {10.1111/j.1365-2966.2009.15054.x}, \href
  {https://ui.adsabs.harvard.edu/abs/2009MNRAS.397.1599Q} {397, 1599}

\bibitem[\protect\citeauthoryear{{Ro{\v{s}}kar}, {Debattista}, {Stinson},
  {Quinn}, {Kaufmann}  \& {Wadsley}}{{Ro{\v{s}}kar} et~al.}{2008}]{Roskar2008}
{Ro{\v{s}}kar} R.,  {Debattista} V.~P.,  {Stinson} G.~S.,  {Quinn} T.~R.,
  {Kaufmann} T.,   {Wadsley} J.,  2008, \mn@doi [\apjl] {10.1086/586734}, \href
  {https://ui.adsabs.harvard.edu/abs/2008ApJ...675L..65R} {675, L65}

\bibitem[\protect\citeauthoryear{{Sellwood}}{{Sellwood}}{2014}]{Sellwood2014}
{Sellwood} J.~A.,  2014, \mn@doi [Reviews of Modern Physics]
  {10.1103/RevModPhys.86.1}, \href
  {https://ui.adsabs.harvard.edu/abs/2014RvMP...86....1S} {86, 1}

\bibitem[\protect\citeauthoryear{{Sellwood} \& {Binney}}{{Sellwood} \&
  {Binney}}{2002}]{Sellwood2002}
{Sellwood} J.~A.,  {Binney} J.~J.,  2002, \mn@doi [\mnras]
  {10.1046/j.1365-8711.2002.05806.x}, \href
  {https://ui.adsabs.harvard.edu/abs/2002MNRAS.336..785S} {336, 785}

\bibitem[\protect\citeauthoryear{{Sestito} et~al.,}{{Sestito}
  et~al.}{2021}]{Sestito2021}
{Sestito} F.,  et~al., 2021, \mn@doi [\mnras] {10.1093/mnras/staa3479}, \href
  {https://ui.adsabs.harvard.edu/abs/2021MNRAS.500.3750S} {500, 3750}

\bibitem[\protect\citeauthoryear{{Soderblom}}{{Soderblom}}{2010}]{Soderblom2010}
{Soderblom} D.~R.,  2010, \mn@doi [\araa]
  {10.1146/annurev-astro-081309-130806}, \href
  {https://ui.adsabs.harvard.edu/abs/2010ARA&A..48..581S} {48, 581}

\bibitem[\protect\citeauthoryear{{Solway}, {Sellwood}  \&
  {Sch{\"o}nrich}}{{Solway} et~al.}{2012}]{Solway2012}
{Solway} M.,  {Sellwood} J.~A.,   {Sch{\"o}nrich} R.,  2012, \mn@doi [\mnras]
  {10.1111/j.1365-2966.2012.20712.x}, \href
  {https://ui.adsabs.harvard.edu/abs/2012MNRAS.422.1363S} {422, 1363}

\bibitem[\protect\citeauthoryear{{Spitoni}, {Verma}, {Silva Aguirre}  \&
  {Calura}}{{Spitoni} et~al.}{2020}]{Spitoni2020}
{Spitoni} E.,  {Verma} K.,  {Silva Aguirre} V.,   {Calura} F.,  2020, \mn@doi
  [\aap] {10.1051/0004-6361/201937275}, \href
  {https://ui.adsabs.harvard.edu/abs/2020A&A...635A..58S} {635, A58}

\bibitem[\protect\citeauthoryear{{Steinmetz} et~al.,}{{Steinmetz}
  et~al.}{2006}]{RAVE1}
{Steinmetz} M.,  et~al., 2006, \mn@doi [\aj] {10.1086/506564}, \href
  {https://ui.adsabs.harvard.edu/abs/2006AJ....132.1645S} {132, 1645}

\bibitem[\protect\citeauthoryear{{Steinmetz} et~al.,}{{Steinmetz}
  et~al.}{2020}]{RAVE2}
{Steinmetz} M.,  et~al., 2020, \mn@doi [\aj] {10.3847/1538-3881/ab9ab8}, \href
  {https://ui.adsabs.harvard.edu/abs/2020AJ....160...83S} {160, 83}

\bibitem[\protect\citeauthoryear{{Stinson}, {Seth}, {Katz}, {Wadsley},
  {Governato}  \& {Quinn}}{{Stinson} et~al.}{2006}]{Stinson2006}
{Stinson} G.,  {Seth} A.,  {Katz} N.,  {Wadsley} J.,  {Governato} F.,   {Quinn}
  T.,  2006, \mn@doi [\mnras] {10.1111/j.1365-2966.2006.11097.x}, \href
  {https://ui.adsabs.harvard.edu/abs/2006MNRAS.373.1074S} {373, 1074}

\bibitem[\protect\citeauthoryear{{Stinson} et~al.,}{{Stinson}
  et~al.}{2013}]{Stinson2013}
{Stinson} G.~S.,  et~al., 2013, \mn@doi [\mnras] {10.1093/mnras/stt1600}, \href
  {https://ui.adsabs.harvard.edu/abs/2013MNRAS.436..625S} {436, 625}

\bibitem[\protect\citeauthoryear{{Sun} et~al.,}{{Sun} et~al.}{2020}]{Sun2020}
{Sun} W.~X.,  et~al., 2020, \mn@doi [\apj] {10.3847/1538-4357/abb1b7}, \href
  {https://ui.adsabs.harvard.edu/abs/2020ApJ...903...12S} {903, 12}

\bibitem[\protect\citeauthoryear{{Twarog}}{{Twarog}}{1980}]{Twarog1980}
{Twarog} B.~A.,  1980, \mn@doi [\apj] {10.1086/158460}, \href
  {https://ui.adsabs.harvard.edu/abs/1980ApJ...242..242T} {242, 242}

\bibitem[\protect\citeauthoryear{{Valentini}, {Bressan}, {Borgani}, {Murante},
  {Girardi}  \& {Tornatore}}{{Valentini} et~al.}{2018}]{Valentini2018}
{Valentini} M.,  {Bressan} A.,  {Borgani} S.,  {Murante} G.,  {Girardi} L.,
  {Tornatore} L.,  2018, \mn@doi [\mnras] {10.1093/mnras/sty1896}, \href
  {https://ui.adsabs.harvard.edu/abs/2018MNRAS.480..722V} {480, 722}

\bibitem[\protect\citeauthoryear{{Vera-Ciro}, {D'Onghia}, {Navarro}  \&
  {Abadi}}{{Vera-Ciro} et~al.}{2014}]{VeraCiro2014}
{Vera-Ciro} C.,  {D'Onghia} E.,  {Navarro} J.,   {Abadi} M.,  2014, \mn@doi
  [\apj] {10.1088/0004-637X/794/2/173}, \href
  {https://ui.adsabs.harvard.edu/abs/2014ApJ...794..173V} {794, 173}

\bibitem[\protect\citeauthoryear{{Wadsley}, {Keller}  \& {Quinn}}{{Wadsley}
  et~al.}{2017}]{Wadsley2017}
{Wadsley} J.~W.,  {Keller} B.~W.,   {Quinn} T.~R.,  2017, \mn@doi [\mnras]
  {10.1093/mnras/stx1643}, \href
  {https://ui.adsabs.harvard.edu/abs/2017MNRAS.471.2357W} {471, 2357}

\bibitem[\protect\citeauthoryear{{Wang}, {Dutton}, {Stinson}, {Macci{\`o}},
  {Penzo}, {Kang}, {Keller}  \& {Wadsley}}{{Wang} et~al.}{2015}]{Wang2015}
{Wang} L.,  {Dutton} A.~A.,  {Stinson} G.~S.,  {Macci{\`o}} A.~V.,  {Penzo} C.,
   {Kang} X.,  {Keller} B.~W.,   {Wadsley} J.,  2015, \mn@doi [\mnras]
  {10.1093/mnras/stv1937}, \href
  {https://ui.adsabs.harvard.edu/abs/2015MNRAS.454...83W} {454, 83}

\bibitem[\protect\citeauthoryear{{Yong} et~al.,}{{Yong}
  et~al.}{2016}]{Yong2016}
{Yong} D.,  et~al., 2016, \mn@doi [\mnras] {10.1093/mnras/stw676}, \href
  {https://ui.adsabs.harvard.edu/abs/2016MNRAS.459..487Y} {459, 487}

\bibitem[\protect\citeauthoryear{{Zhang} et~al.,}{{Zhang}
  et~al.}{2021}]{Zhang2021}
{Zhang} M.,  et~al., 2021, arXiv e-prints, \href
  {https://ui.adsabs.harvard.edu/abs/2021arXiv210900746Z} {p. arXiv:2109.00746}

\makeatother
\end{thebibliography}




\appendix \label{sec:append}
\renewcommand{\thesubsection}{\Alph{subsection}}
\counterwithin{figure}{subsection}
\counterwithin{table}{subsection}
\subsection{Additional figures}
    
Figure~\ref{fig:fehz0_hilow} shows the same figure as Figure~\ref{fig:feagez0} but separated by the high and \lowalpha~disks.
It is clear that the major turnover points only exist in the \lowalpha~disk, as expected, since the satellite infall that causes the turnover points happened after the \lowalpha~has formed. 

\begin{figure*}[htp]
    \centering
    \includegraphics[width=\textwidth]{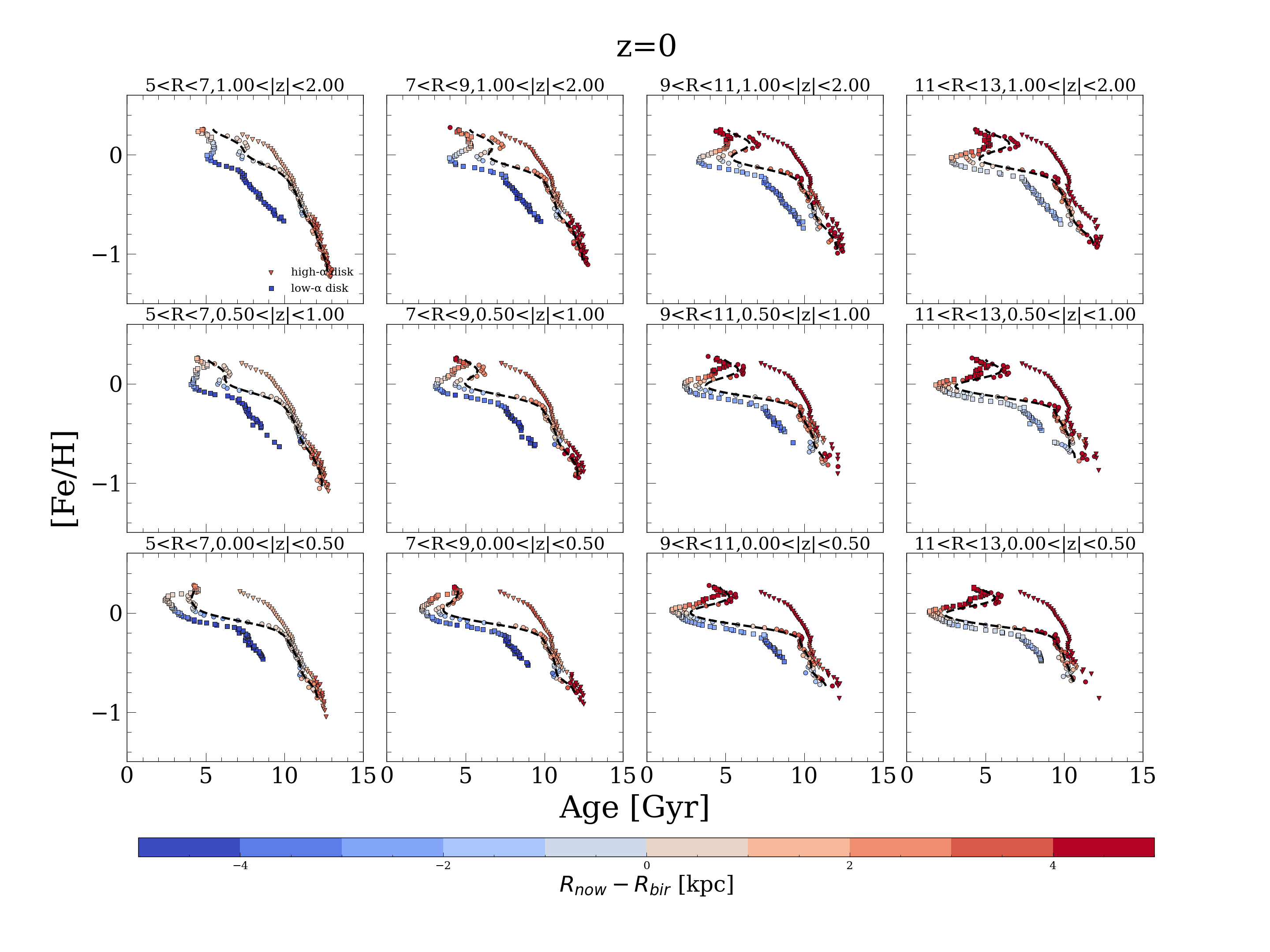}
    \caption{Same as Figure~\ref{fig:feagez0} but separated into the high- and \lowalpha disks. The square markers show the \lowalpha disk stars, the triangles show the \highalpha\ disk stars, and the circles show all stars together. }
    \label{fig:fehz0_hilow}
\end{figure*}

Figure~\ref{fig:fehz0_nomig} shows the stars that are born and remain within each spatial bin of Figure~\ref{fig:feagez0}, thereby illustrating the AMR in the absence of migration. Note the AMR turning points are all still present, as the migration does not create these. 

\begin{figure*}[htp]
    \centering
    \includegraphics[width=\textwidth]{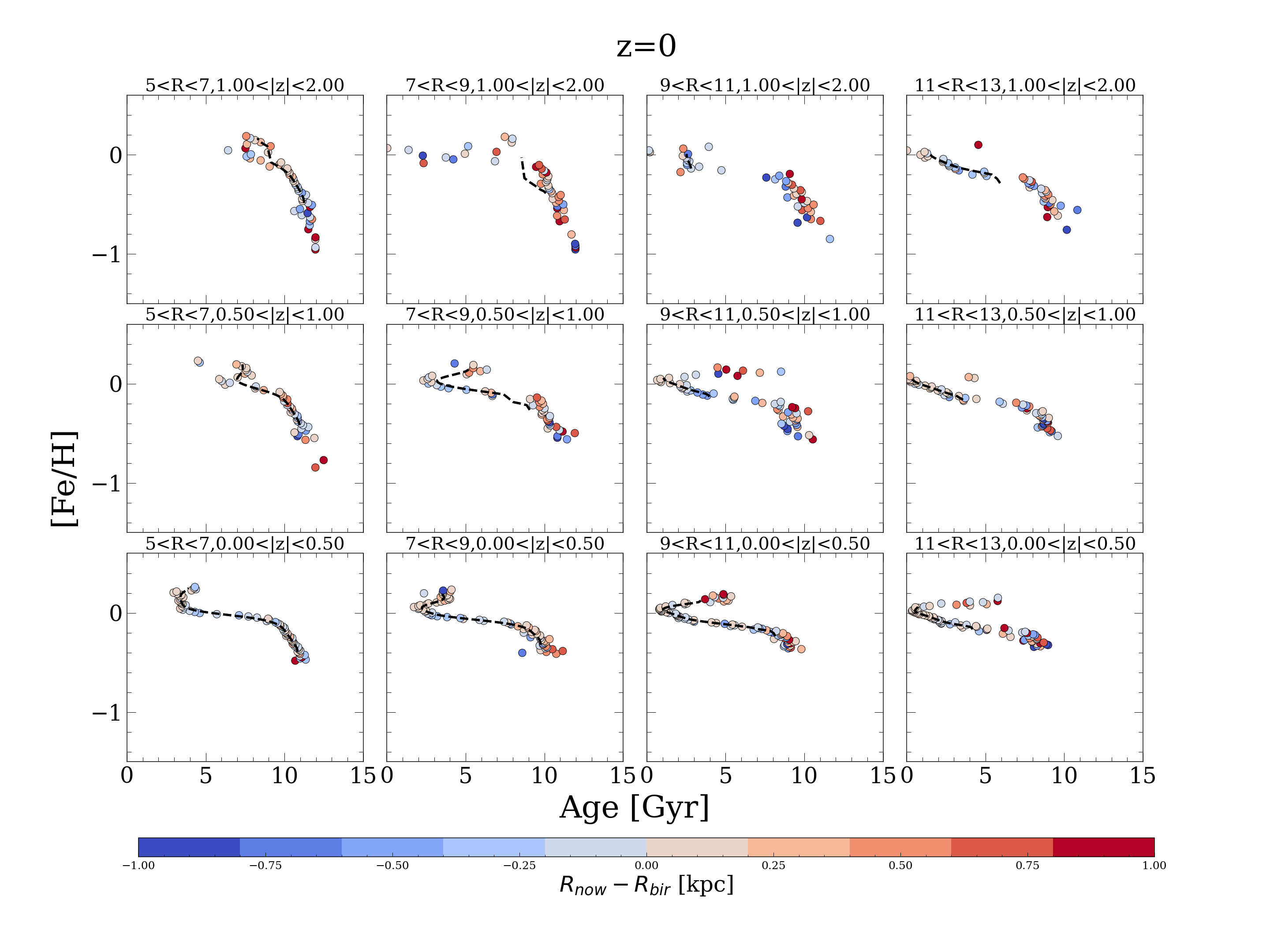}
    \caption{Same as Figure~\ref{fig:feagez0} but only for stars that are born within each spatial bin. 
    It is clear that without radial migration, the turnover points still exist in most bins. 
    However, the most metal-enriched and metal-poor part of the current day AMR (see Figure~\ref{fig:feagez0}) migrated from other spatial bins as expected. }
    \label{fig:fehz0_nomig}
\end{figure*}

Figure~\ref{fig:fehsnap} shows the metallicity distribution of stars born between each snapshot around the merger in the radial bin of R = 7 $-$ 9 kpc with galactic height $|z| <$ 0.5 kpc. 
For each snapshot, we calculated the mean (points) and standard deviation (errorbars) in [Fe/H] for the newly born stars.
The shaded area in gray marks the approximate time of the merger (where Age=0 is the present day at redshit 0). It is clear that when the merger happened, stars started to form at a lower metallicity bin as the infall brought in metal-poor gas (also shown in the dip in ISM metallicity in Figure~\ref{fig:feage_ism} around 4 Gyr), causing the turnover point.

\begin{figure*}[htp]
    \centering
    \includegraphics[width=\textwidth]{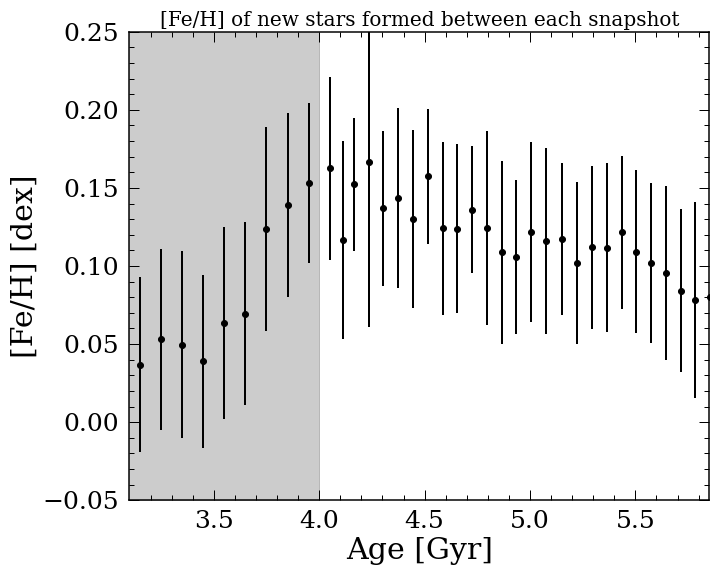}
    \caption{The mean (points) and standard deviation (errorbars) of metallicity of stars born between each snapshot ($\sim$ 200 Myr apart) around the satellite infall in the radial bin of 7-9 kpc with galactic height $|z| <$ 0.5 kpc.
    The gray area marked the approximate time of the satellite infall (first passage around 4 Gyr). 
    Combined with Figure~\ref{fig:feage_ism}, we can see the decrease in mean metallicity of stars born after the merger, creating the turnover point.}
    \label{fig:fehsnap}
\end{figure*}

Figure~\ref{fig:migrationhist} shows the migration strength in different age bin for the stars in the simulation (black dots) and the model in \cite{Frankel2019} (3.9 kpc $\sqrt{\tau/7 Gyr}$; black line) for disk stars (selected by choosing only stars with radius $>$ 5 kpc). 
We calculated the strength ($\sigma$) by taking the standard deviation of \dr~in each age bin for $R > $ 5 kpc to exclude the bulge stars.
The shape of the migration strength in the simulation matches with that from \cite{Frankel2019} until the most recent merger with a relatively constant offset.

\begin{figure*}[htp]
    \centering
    \includegraphics[width=0.5\textwidth]{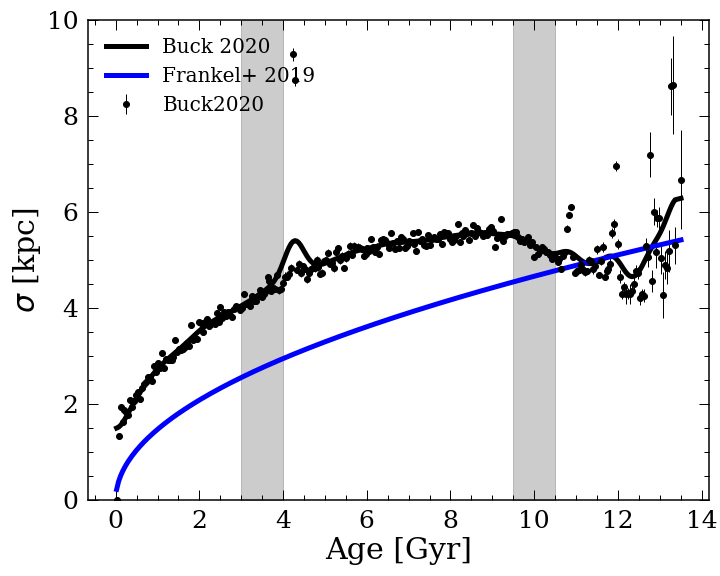}
    \caption{The radial migration strength for disk stars as function of age.
    The black dots show the radial migration strength in different age bin for the simulation (with $R >$ 5 kpc), the black line shows the smoothed curve over the black dots, and the blue line shows the migration strength for the MW as inferred in Frankel et al. (2019) (3.9 kpc $\sqrt{\tau/7 Gyr}$). The grey shaded area shows the approximate time of the two most recent merger events ($\sim$ 3 and 10 Gyrs ago). }
    \label{fig:migrationhist}
\end{figure*}


\bsp	
\label{lastpage}
\end{document}